\documentclass[ prb, aps, superscriptaddress, twocolumn]{revtex4}
\usepackage{mathrsfs}
\usepackage{array}
\usepackage{amsmath}
\usepackage{graphicx}
\usepackage{amstext}
\usepackage{amsfonts}
\usepackage{bm}
\usepackage{epstopdf}

\begin{document}
\title{Exact master equation and non-Markovian decoherence dynamics of \\ Majorana zero modes under gate-induced charge fluctuations}
\author{Hon-Lam Lai}
\affiliation{Department of Physics and Center for Quantum
information Science, National Cheng Kung University, Tainan 70101,
Taiwan}
\author{Pei-Yun Yang}
\affiliation{Department of Physics and Center for Quantum
information Science, National Cheng Kung University, Tainan 70101,
Taiwan}
\affiliation{Physics Division, National Center for Theoretical Sciences,
National Tsing Hua University, Hsinchu 30013, Taiwan}
\author{Yu-Wei Huang}
\affiliation{Department of Physics and Center for Quantum
information Science, National Cheng Kung University, Tainan 70101,
Taiwan}
\author{Wei-Min Zhang}
\email{wzhang@mail.ncku.edu.tw}
\affiliation{Department of Physics
and Center for Quantum information Science, National Cheng Kung
University, Tainan 70101, Taiwan}
\begin{abstract}
In this paper, we use the exact master equation approach to investigate the decoherence dynamics of Majorana zero modes in the Kitaev model, a 1D $p\,$-wave spinless topological superconducting chain (TSC), that is disturbed by gate-induced charge fluctuations. The exact master equation is derived by extending Feynman-Vernon influence functional technique to fermionic open systems involving pairing excitations. We obtain the exact master equation for the zero-energy Bogoliubov quasiparticle (bogoliubon) in the TSC, and then transfer it into the master equation for the Majorana zero modes. Within this exact master equation formalism, we can describe in detail the non-Markovian decoherence dynamics of the zero-energy bogoliubon as well as Majorana zero modes under local perturbations. We find that at zero temperature, local charge fluctuations induce level broadening to one of the Majorana zero modes but there is an isolated peak (localized bound state) located at zero energy that partially protects the Majorana zero mode from decoherence. At finite temperatures, the zero-energy localized bound state does not precisely exist, but the coherence of the Majorana zero mode can still be partially but weakly protected, due to the sharp dip of the spectral density near the zero frequency. The decoherence will be enhanced as one increases the charge fluctuations and/or the temperature of the gate.
\end{abstract}

\maketitle
\section{Introduction}
Realization of quantum computation could provide a tremendous revolution in sciences and technologies [\onlinecite{Nielsen and Chuang}], yet there are many obstacles for practical quantum information processing. One of the main difficulties is the decoherence problem, i.e., quantum coherence loss due to the quantum entanglement between the system and its environment. Topological quantum computation (TQC) is thought to be robust against decoherence. It realizes quantum computation in a way that is topologically protected from local perturbations [\onlinecite{Kitaev2003}-\onlinecite{PRB2017}]. This is mainly due to topological properties of the system involving zero-energy excitations. These zero-energy excitations, which are known as Majorana zero modes, are exotic particles bound by topological defects or vortices of the system. These Majorana zero modes lead to the system ground state degeneracy. Qubits are defined by these degenerate ground states, which are separated from low-lying excited states of the system by a finite energy gap $\Delta$. Unitary operations of qubits can be done by braiding of Majorana zero modes. These operations are considered to be determined only by the topology of braiding trajectories, not sensitive to the detail of those trajectories so that qubit operations are immune to local perturbations induced by the environment [\onlinecite{Kitaev2003}].

The realization of such a TQC system requires a physical system in a topological phase that exhibits non-Abelian statistics. Kitaev first proposed such a TQC by a toy model of a 1D p-wave spinless superconducting chain [\onlinecite{Kitaev2001}]. He showed that, in the topological phase, there are two Majorana zero modes localized at the ends of the chain that can be considered as the building blocks for topological quantum computers. Since then, several proposals based on proximity effect of interfaces between superconductors and topological insulators have been considered [\onlinecite{Kitaev2001}-\onlinecite{Nature2016}]. In particular, 1D semiconducting wires with spin-orbit coupling deposited in a $s$-wave superconductor can be engineered to realize Kitaev's model [\onlinecite{PRL2010c},\onlinecite{PRL2010d}] and perform TQC operations [\onlinecite{Nature2011}].

However, the robustness against decoherence of such TQC systems, based on Majorana zero modes, has been questioned. It has been argued that the protection from decoherence for 
Majorana zero modes will be broken down with certain kinds of operations in realistic situation  [\onlinecite{Goldstein2011}-\onlinecite{PRB2013}]. For example, Budich \emph{et al.} [\onlinecite{Budich2012}] considered the problem from an open system perspective. They argued that the protection against decoherence based on Majorana zero modes will fail if these zero modes are coupled to an ungapped fermionic bath. This is because single electrons can tunnel between the topological superconducting chain (TSC) and the bath without paying any energy, leading to decoherence. Another work is done by Goldstein and Chamon [\onlinecite{Goldstein2011}], in which they considered the coupling between a gapped fermionic bath and Majorana zero modes through a bosonic field. By calculating perturbatively two-time correlation functions, they showed that even if both the system and the environment are gapped, certain kind of bosonic field fluctuations can still damage Majorana zero modes. In contrast, it was pointed out that Majorana zero mode decoherence can be well suppressed if temperature of the environment is low enough compared to the superconducting energy gap [\onlinecite{Goldstein2011},\onlinecite{PRB2012}]. In view of the model given by Goldstein and Chamon, Schmidt \emph{et al}. [\onlinecite{Daniel2012}] suggested a more realistic model based on an experimental setup [\onlinecite{Nature2011}], in which the braiding of Majorana zero modes are performed by manipulating the superconducting chain through tunable external gates. As a result, Majorana zero modes are affected by gate-induced charge fluctuations. They have studied the life time of the zero-energy bogoliubon as a function of temperature in the low superconducting energy gap limit using Fermi's golden rule, and obtained an optimal region for topological states against decoherence [\onlinecite{Daniel2012}].

However, these investigations on Majorana zero modes decoherence discussed above are carried out with certain approximations, and are mainly valid for Markov processes. Because TQC systems are realized through various solid-state engineering manners, the decoherence dynamics of TQC systems in realistic situations involves most likely non-Markovian processes. Therefore, a systematic treatment of non-Markovian dynamics in open quantum systems is desired for a better understanding of
topological state decoherence dynamics. In the present work, we treat the TSC system with gate-induced charge fluctuations investigated in Ref.~[\onlinecite{Daniel2012}] as an open quantum system. We then derive exactly the master equation of the system, using the approach we developed recently [\onlinecite{PRB2008, NJP2010,ANNP2012,PRL2012,PRB2015}] through the Feynman-Vernon influence functional technique [\onlinecite{Fey1963}]. In this formalism, the environment-induced dissipation and fluctuation decoherence
dynamics to Majorana zero modes are fully determined by Schwinger-Keldysh's nonequilibrium Green functions [\onlinecite{Schwinger1961,Keldysh1965,Kadanoff1962}], which depict all possible non-Markovian memory dynamics. Explicitly, we find that at zero temperature, charge fluctuations induce partial decoherence to Majorana zero modes, due to the existence of a localized bound state which is dissipationless as a long-time non-Markovian dynamics that we discovered recently [\onlinecite{PRL2012}]. At a finite temperature, the localized bound state is not formed, but coherence of Majorana zero modes can still be partially but weakly protected, due to the sharp dip of the spectral density near the zero frequency for charge fluctuations. In both zero and finite temperatures, short-time non-Markovian memory effect [\onlinecite{PRA2015}]
shows up in the strong charge fluctuation regime.
Under weak charge fluctuations, by simply taking a second-order perturbation approximation from our exact formalism, we reproduce the Markovian decoherence result obtained by Schmidt \emph{et al}.~[\onlinecite{Daniel2012}]. We also show explicitly from two-time correlations of Majorana zero modes that local perturbations may disturb certain Majorana qubit operations, due to both Markovian and non-Markovian decoherence effects.

The rest of the paper is organized as follows. In Section II, we briefly describe the Kitaev model of a 1D $p\,$-wave spinless superconducting chain under the disturbance of gate-induced charge fluctuations that was introduced in [\onlinecite{Daniel2012}]. We then extend the Feynman-Vernon influence functional approach to the fermionic coherent state representation with pairing excitations, to derive the exact master equation that can fully describe the decoherence dynamics of the zero-energy bogoliubon. By re-expressing the bogoliubon operators in terms of Majorana zero-mode operators, we also obtain the master equation for Majorana zero modes, and it shows that local perturbations only affect on one of the two Majorana zero modes. Also, the resulting master equation turns out to be extremely simple that corresponds to the master equation of a spin-like pure dephasing model coupled to a fermionic bath. In Sec.~III, we analyze in detail the decoherence dynamics of Majorana zero modes under local perturbations at zero and finite gate temperatures.
A comparison between the non-Markovian dynamics obtained in the exact decoherence solution and the Markovian dynamics obtained from the second-order perturbation theory is also presented.
In Sec.~IV, we investigate further two-time correlation functions of the left and right Majorana zero modes. The results show that the topologically non-local correlations between the left and right Majorana zero modes can be disturbed by local perturbations. This indicates that some Majorana qubit operations are not really immune to local perturbations. Finally, a conclusion is given in Sec.~V. The appendices contain detailed derivations of the formulae.

\section{model and formulation}
\subsection{The TSC model with charge fluctuations}
In order to study decoherence dynamics of Majorana zero modes, we begin with the model introduced by Schmidt {\it et al.} [\onlinecite{Daniel2012}]. The model Hamiltonian (with the unit $\hbar=1$) is
\begin{align}
H=& \, H_{TSC}+H_G+H_C \notag\\
=&\!\sum_{i} \! \big[\!-\!\frac{\Delta}{2}c_{i+1}c_i \!-\!\frac{w}{2}c^\dag_{i+1}c_i \!+\!H.c. \!-\!\mu_ic^\dag_ic_i \big] \notag \\
& + \sum_{p}\epsilon_pc^\dag_pc_p  + \eta\delta Q\sum_{i}F_ic^\dag_ic_i .   \label{tH}
\end{align}
The first summation is the TSC Hamiltonian which is described by an one-dimensional tight-binding chain with $N$ sites
[\onlinecite{Kitaev2001}], and $c^\dag_i$ ($c_i$) is the electron creation (annihilation) operator at site $i$, $\Delta$ is the p-wave superconducting gap, $w$ is the hopping amplitude and $\mu_i$ is the electron chemical potential at site $i$.
The TSC can be tuned by a controlling gate which is modeled as non-interacting electron gases given by the second summation in Eq.~(\ref{tH}),
where $c^\dag_p$ ($c_p$) is the electron creation (annihilation) operator in the gate with energy $\epsilon_p$. The charge fluctuations of the controlling gate induce quasiparticle excitations in the TSC. It is described by the coupling Hamiltonian between the TSC and the gate, the last summation in Eq.~(\ref{tH}),
where $\eta$ is a coupling parameter, $\delta Q$ is the charge fluctuation of the gate and $F_i$ is a profile function which limits the range of the chain affected by the gate. Here, the parameter $F_i$ equals to unity for small $i$ and zero for large $i$. As pointed out in Ref.~[\onlinecite{Daniel2012}], it is not important how exactly large $i$ is for $F_i$ to be zero.

\begin{figure}
\centerline{\scalebox{0.2}{\includegraphics{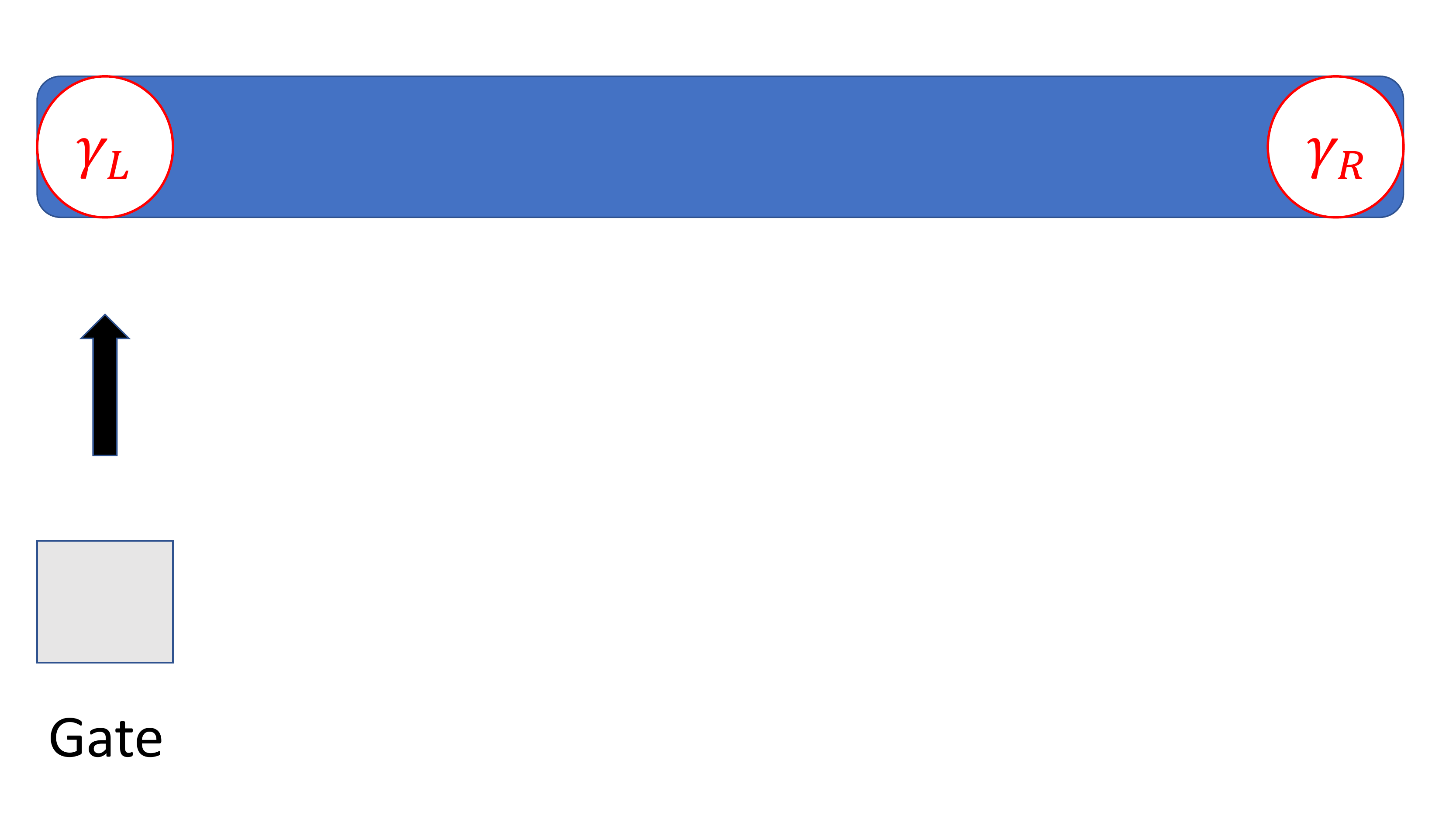}}}
\caption{A schematic diagram of two Majorana fermions localized at the two ends of the TSC in the Kitaev model, and can be tuned by a controlling gate coupled to one of the ends.}\label{system}  \label{fig1}
\end{figure}

By a Bogoliubov transformation to the TSC Hamiltonian and setting $\mu_i=0$ for simplicity, $H_{TSC}$ can be rewritten in terms of Bogoliubov quasiparticles or simply bogoliubons [\onlinecite{Daniel2012}],
\begin{align}
H_{TSC}=\sum_{k\neq 0}E_kb^\dag_kb_k ,
\label{TSC2}
\end{align}
where $E_k=\sqrt{w^2\cos^2 (k/2)+\Delta^2 \sin^2 (k/2)}$
is the quasiparticle excitation energy with $k=2\pi m/(N+1)$, $m=1,2,...,(N-1)/2$.
The zero-energy bogoliubon ($b^\dag_0, b_0$), which is absent in Hamiltonian~(\ref{TSC2}),
consists of two Majorana zero modes located at the left and right ends of the chain
[\onlinecite{Kitaev2001}], see Fig.~\ref{fig1}. The relations between bogoliubons $b_0,b_k$ and
the original electron operator $c_i$'s have been derived explicitly in Ref.~[\onlinecite{Daniel2012}].
The two Majorana zero modes can be described by Majorana operators $\gamma_L$ and $\gamma_R$ that are defined as
\begin{align}\label{gamma}
\gamma_L=-i(b_0-b^{\dag}_0)~,~~~\gamma_R=b_0+b^{\dag}_0.
\end{align}
It is obviously that $\gamma_L^\dag=\gamma_L, \gamma_R^\dag=\gamma_R$, and they obey the anticommutation relations:
$\{ \gamma_i~,~ \gamma_j\}= 2\delta_{ij};~ i, j:= L, R$,
which indicate that Majorana zero modes exhibit non-Abelian statistics.
It also shows that the TSC ground state is two-fold degenerate, i.e., the degenerate states of the zero-energy bogoliubon with particle number $b^\dag_0b_0=0~\rm{and}~1$, respectively, made by Majorana zero modes.

Based on this model, Schmidt \emph{et al}.~[\onlinecite{Daniel2012}] have investigated the decoherence of
the zero-energy bogoliubon induced by charge fluctuation $\delta Q$. Explicitly,
they rewrite $H_C$ in the bogoliubon representation as
\begin{align}
H_C=\eta\delta Q\sum_{k,k'}[b^\dag_k X_{kk'} b_{k'}+\frac{1}{2}(b_k Y_{kk'} b_{k'}+H.c.)] ,
\label{HI2}
\end{align}
where the summation includes the zero-energy as well as all the finite-energy bogoliubons. The coefficient $X_{kk'}$ is a hopping amplitude from the mode $k'$ to $k$, while the coefficient $Y_{kk'}$ is a pairing excitation amplitude between modes $k$ and $k'$. The zero-energy bogoliubon decoherence is induced by the excitations from zero-energy bogoliubon to other finite-energy bogoliubon states, due to charge fluctuations. Thus, one can only keep the terms with either $k$ or $k'$ to be zero in Eq.~(\ref{HI2}). It has further been shown [\onlinecite{Daniel2012}] that
\begin{align}
X_{0k}=X_{k0}&=Y_{0k}=-Y_{k0}\notag\\
&=-\frac{\sqrt{1-\delta^2}}{\sqrt{N+1}}
\frac{\sin(k+\alpha_k)-\delta\sin(\alpha_k)}{1+\delta^2-2\delta\cos(k)} ,
\label{XY}
\end{align}
where $\delta=(\Delta-w)/(\Delta+w)$ and
\begin{align}
\alpha_k=-\arccos\Bigg(\frac{w\cos^2(k/2)+\Delta\sin^2(k/2)}{\sqrt{w^2\cos^2(k/2)+\Delta^2\sin^2(k/2)}}\Bigg).
\label{alphak}
\end{align}

From the above considerations, Schmidt \emph{et al}.~[\onlinecite{Daniel2012}] computed the life time of the zero-energy bogoliubon using Fermi's golden rule and studied the reliable parameter regime that could provide a better protection of the zero-energy bogoliubon from decoherence. Fermi's golden rule is based on the second-order perturbation theory with respect to the coupling strength $\eta$. In this paper, we shall study the decoherence dynamics of Majorana zero modes, using the exact master equation for the zero-energy bogoliubon as well as the two Majorana zero modes that we will derive in the next subsection. This could provide a complete understanding of the decoherence dynamics of the Majorana zero modes, and also the real-time correlations between them.

\subsection{Exact master equation of the zero-energy bogoliubon}
To describe decoherence dynamics of the zero-energy bogoliubon, we treat the zero-energy bogoliubon as the system of interest. The other finite-energy bogoliubons and the controlling gate together are considered as its environment. Put all the considerations in the last subsection together, the total Hamiltonian Eq.~(\ref{tH}) can be written as
\begin{align}
H=&\sum_{k\neq 0}E_kb^\dag_kb_k+\sum_{p}\epsilon_pc^\dag_pc_p\notag\\
&+\eta\delta Q\sum_{k\neq 0}X_{0k}[b^\dag_0b_{k}+b^\dag_0b^\dag_{k}+H.c.].
\label{HT}
\end{align}
The decoherence dynamics of the zero-energy bogoliubon is then determined by the reduced density matrix $\rho(t)$ of the system.
It is obtained by tracing out completely the degrees of freedom of the environment, including all the
non-zero bogoliubon k-modes in the TSC and all the degrees of freedom in the controlling gate,
\begin{align}
\rho(t)={\rm Tr}_{E}[U(t,t_0)\rho_{\rm tot}(t_0)U^\dag(t,t_0)] .
\label{trace rho}
\end{align}
Here $U(t,t_0)=\exp[-iH(t-t_0)]$ is the time-evolution operator of the total system, and $\rho_{\rm tot}(t_0)$ is the total density matrix at the initial time $t_0$.
In the Bogoliubov quasiparticle picture, the superconducting state is the vacuum state of bogoliubon operators. Thus the initial state of the total system can be setup as
\begin{align}
\rho_{tot}(t_0)=\rho(t_0)\otimes\rho_B(t_0)\otimes\rho_G(t_0).
\label{rho initial}
\end{align}
Here $\rho(t_0)$ can be an arbitrary initial state of the zero-energy bogoliubon, $\rho_B(t_0)=|\Omega\rangle\langle \Omega|$ with $|\Omega\rangle$ being the BCS state, i.e. the vacuum state of all the finite-energy bogoliubons: $b_k|\Omega\rangle=0$ (the subscript $k \neq 0$ means that it does not include the zero-energy bogoliubon), and $\rho_G(t_0)=\frac{1}{Z}e^{-\beta (H_G-E_F N_G)}$ is the thermal equilibrium state of the controlling gate with initial temperature $\beta=1/k_BT$ and particle number operator $N_G=\sum_pc^\dag_p c_p$. Because the controlling gate applies to the TSC, usually we require that this gate temperature should be lower than the critical temperature of the corresponding superconductor.

The master equation for $\rho(t)$  can be derived rigorously using the path integral approach in the fermionic coherent state representation that we have developed in our previous works [\onlinecite{PRB2008}-\onlinecite{PRB2015}]. However, it is different from our previous derivation, here the coupling Hamiltonian between the system (the zero-energy bogoliubon) and the environment contains fermionic pair creation and annihilation terms [see the last term in Eq.~(\ref{HT})]. This makes the derivation of the master equation much more difficult, in comparison to what we have done in Refs.~[\onlinecite{PRB2008}-\onlinecite{PRB2015}]. The detailed derivation is given in Appendix A. The resulting exact master equation for the reduced density matrix of the zero-energy bogoliubon is
\begin{align}
\frac{d}{dt}\rho(t)& =-i[E'_0(t,t_0)b^\dag_0b_0,\rho(t)] \notag  \\
&+ \! \lambda (t,t_0)\{2b_0\rho (t)b^\dag_0\!-\!b^\dag_0b_0\rho (t)\!-\!\rho (t)b^\dag_0b_0\}  \notag \\
&+\! \tilde{\lambda}(t,t_0)\{b^\dag_0\rho (t)b_0\!+\!b^\dag_0b_0\rho (t)\!-\!b_0\rho (t)b^\dag_0\!-\!\rho (t)b_0b^\dag_0\} \notag \\
&+\! \Lambda (t,t_0)b^\dag_0\rho (t)b^\dag_0\!+\!\Lambda^*(t,t_0)b_0\rho (t)b_0,
\label{ME}
\end{align}
where $E'_0(t,t_0)$ is the renormalized energy of the zero-energy bogoliubon. As we will show in the end of this section, this renormalized energy indeed equals to zero so that the zero-energy bogoliubon remains to have zero energy after the charge fluctuation influence is included. The coefficients $\lambda(t,t_0)$ and $\tilde{\lambda}(t,t_0)$ in the master equation describe respectively the dissipation and fluctuation dynamics of the zero-energy bogoliubon, which are the typical environment-induced decoherence effects due to the coupling between the zero and finite-energy bogoliubons. Moreover, $\Lambda (t,t_0)$ is a new dissipation coefficient induced by pairing excitations between the zero and the finite-energy bogoliubons, also coming from charge fluctuations. This pairing associated dissipation is a new enter that does not occur in the exact master equation in our previous formalism where pairing excitations were not included [\onlinecite{PRB2008}-\onlinecite{PRB2015}].

In the exact master equation (\ref{ME}), all the time-dependent dissipation and fluctuation coefficients are determined by
Schwinger-Keldysh's nonequilibrium Green functions through the following relations
\begin{subequations}
\label{Coef}
\begin{align}
& \begin{pmatrix}
 iE'_0 (t,t_0)+\lambda (t,t_0) & \Lambda (t,t_0) \\
\Lambda^*(t,t_0) &\!\!\!\! iE'_0 (t,t_0)+\lambda (t,t_0)
\end{pmatrix} \notag  \\
&~~~~~~~~~~~~~~~~~~~~~~~~~~~~~~=-\dot{\bm{U}}(t,t_0)\bm{U}^{-1}(t,t_0),   \label{re_energy} \\
&\widetilde{\lambda}(t,t_0)=\dot{\bm{V}}_{11}(t,t)\!-\![\dot{\bm{U}}(t,t_0)\bm{U}^{-1}(t,t_0)\bm{V}(t,t)\!+\!H.c.]_{11} ,
\end{align}
\end{subequations}
where  $\bm{U}(t,t_0)$ and $\bm{V}(t,t)$ are $2 \times 2$ matrix functions which are the generalized non-equilibrium Green functions
incorporating paring excitations. These Green functions satisfy the integro-differential equations,
\begin{subequations}
\label{UV}
\begin{align}
&\frac{d}{dt}\bm{U}(t,t_0)+ \!\!\int^t_{t_0} \!\!\! d\tau \tilde{\bm{G}}(t,\tau)\bm{U}(\tau,t_0)\!=\!0\label{UVa},  \\
&\bm{V}(\tau,t) =\! \int_{t_0}^{\tau} \!\!\!\! d\tau' \!\! \int_{t_0}^t \!\!\! d\tau{''}\bm{U}(\tau,\tau')\bm{G}(\tau'',\tau')\bm{U}^\dag(t,\tau{''}) , \label{UVb}
\end{align}
\end{subequations}
subjected to the initial conditions $\bm{U}(t_0,t_0)=\bm{I}$.

Non-Markovian dynamics of open quantum systems is depicted by the above nonequilibrium  Green functions [\onlinecite{PRL2012}] through the integral memory kernels $\tilde{\bm{G}}(t,\tau)$ and $\bm{G}(t,\tau)$,
\begin{align}
&\tilde{\bm{G}}(t,\tau)=
2{\rm Re}[\bm{G}(t,\tau)] ~,
~ \bm{G}(t,\tau)=
g(t,\tau)
\begin{pmatrix}
1 & -1\\
-1 & 1
\end{pmatrix},
\label{Gtilde}
\end{align}
where $g(t,\tau)$ is a two-time correlation function between the system and the environment,
\begin{align}
g(t,\tau)=\eta^2 \!\! \int_{-\infty}^{\infty} \!\!\!\! d\omega J_C(\omega)\langle \delta Q(t)\delta Q(\tau)\rangle_G e^{-i\omega(t-\tau)}.
\label{g}
\end{align}
In Eq.~(\ref{g}), $J_C(\omega)=\sum_{k\neq 0}X_{0k}^2\delta(\omega-E_k)$ is the spectral density of the non-zero bogoliubon modes of the superconducting chain. The charge fluctuation $\langle \delta Q(t)\delta Q(\tau)\rangle_G$ induced from the controlling gate is determined by the correlation function [\onlinecite{Daniel2012}]
\begin{align}
\langle \delta Q(t)\delta Q(\tau)\rangle_G=\int_{-\infty}^{\infty} \!\!\!\! d\omega J_G(\omega)e^{-i\omega(t-\tau)} , \label{cfc}
\end{align}
with the spectral density
\begin{align}
J_G(\omega)=B_Ge^{-\omega^2/8E_F \omega_c}\frac{\omega}{1-e^{-\omega/k_BT}},
\label{JG}
\end{align}
where $E_F$ is the Fermi energy of the controlling gate, $T$ is the temperature of the gate, $\omega_c$ is a cut-off frequency of charge fluctuations, and $B_G$ is a factor determined by the structure of the gate [\onlinecite{Daniel2012}].

\begin{figure}
\centerline{\scalebox{0.45}{\includegraphics{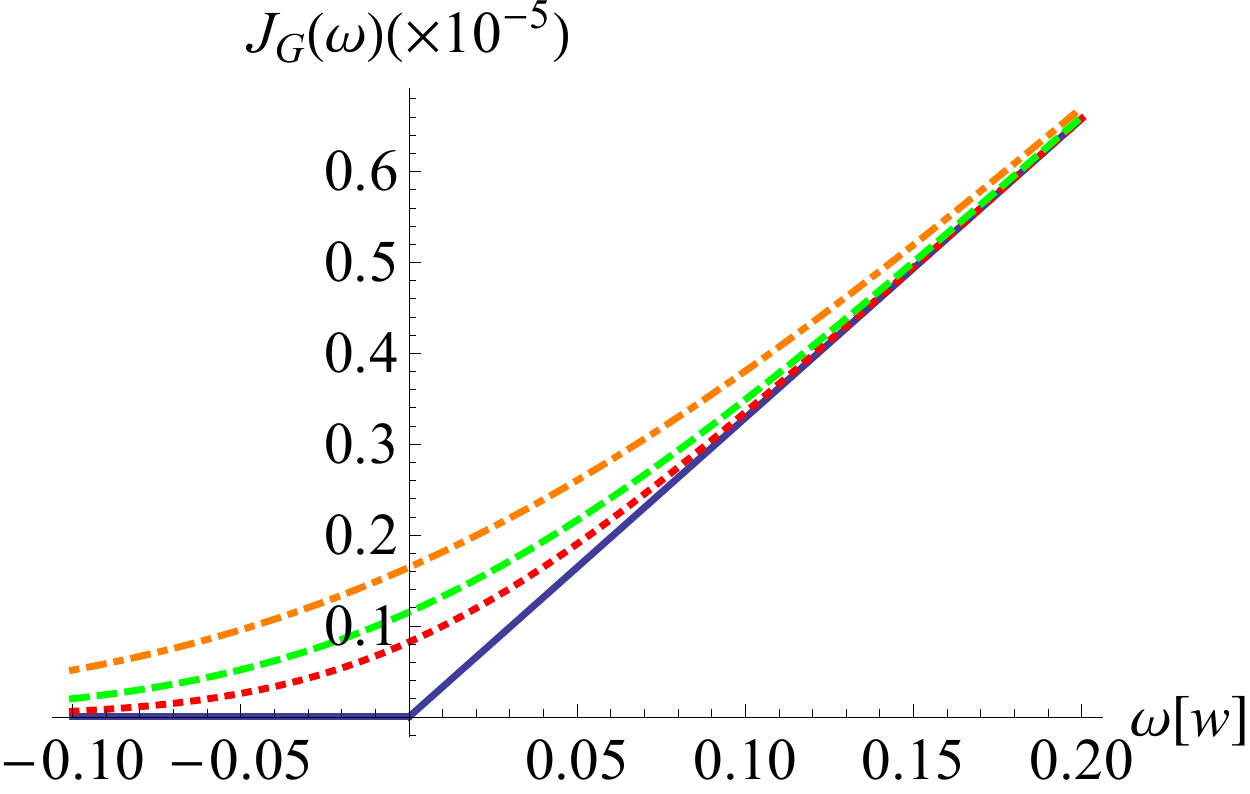}}}
\caption{(color online) The 1D gate spectral density $J_G(\omega)$ at different temperatures $k_BT=0$ (solid), $0.025w$ (dotted), $0.035w$ (dashed), $0.05w$ (dotted-dashed). The Fermi energy $E_F=100w$ and the cut-off frequency $\omega_c=5w$, $B_G=1/(8E_F\omega_c)$. Here we take all parameters in terms of the unit of the hopping amplitude $w$.} \label{Gspec}
\end{figure}

Notice that $J_G(-|\omega|)=0$ at zero temperature (see Fig.~\ref{Gspec}, and also Ref.~[\onlinecite{Daniel2012}]), the negative frequency of $J_G(\omega)$ is dominated only by thermal fluctuations. Moreover, we can rewrite Eq.~(\ref{g}) in a more compact form
\begin{align}
g(t, \tau)=g(t-\tau)=\int_{-\infty}^{\infty}d\omega J(\omega)e^{-i\omega(t-\tau)},  \label{singg}
\end{align}
where the total spectral density $J(\omega)$ is given by
\begin{align}
&J(\omega)=\eta^2\int_{-\infty}^{\infty}d\omega' J_G(\omega-\omega')J_C(\omega')\notag\\
&=\eta^2B_G\sum_{k\neq 0}X_{0k}^2\exp\!\Big[\!\!-\!\!\frac{(\omega-E_k)^2}{8E_F \omega_c}\Big]\frac{\omega-E_k}{1-e^{-(\omega-E_k)/k_BT}},
\label{JT}
\end{align}
in which $X_{0k}$ and $E_k$ are given by Eq.~(\ref{XY}) and that after Eq.~(\ref{TSC2}), the factor $\eta^2B_G$ characterizes the
strength of charge fluctuations.
It is easy to show that all the matrix elements of the integral kernel $\tilde{\bm{G}}(t,\tau)$ are real numbers so that all the matrix elements of the Green function $\bm{U}(t,t_0)$ [see Eq.~(\ref{UVa})] are also real. As a result, $\Lambda^*(t)=\Lambda(t)$, and the renormalized energy $E'_0(t)$ [Eq.~(\ref{re_energy})], which is proportional to the imaginary part of $[\dot{\bm{U}}(t,t_0)\bm{U}^{-1}(t,t_0)]_{11}$, must be zero as we pointed out earlier. This indicates that the renormalization from charge fluctuations does not shift the energy position of the zero-energy bogoliubon.  This can be regarded as one of the topological features of zero-energy excitations in topological systems.
However, the decoherence dynamics of zero-energy bogoliubon (and Majorana zero modes) is much more complicated than the energy renormalization, as we will study in the next Section.

\subsection{Exact master equation for Majorana zero modes}
The master equation (\ref{ME}) given in the above subsection can be used to fully describe the decoherence dynamics of the zero-energy bogoliubon. We can also deduce Eq.~(\ref{ME}) into the master equation for the Majorana operators $\gamma_L$ and $\gamma_R$. In fact, we can rewrite the last term of Eq.~(\ref{HT}) as
\begin{align}
H_C=-i\eta\delta Q\sum_{k\neq0}X_{0k}\gamma_L(b_k+b^\dag_k) .
\label{HI3}
\end{align}
It shows that the right Majorana mode $\gamma_R=b_0+b^\dag_0$ does not couple to the environment (i.e.~it does not couple to the finite-energy bogoliubons through the controlling gate). Only the left Majorana mode $\gamma_L$ is affected by the charge fluctuations. In other words, Eq.~(\ref{HI3}) describes a typical local perturbation to the Majorana zero modes, as illustrated in Fig.~\ref{fig1}.

To see explicitly the decoherence dynamics of Majorana zero modes (not only the decoherence of the zero-energy bogoliubon),
we find that the nonequilibrium Green function of the left and right Majorana modes can be expressed in terms of the matrix elements $u_{ij}(t,t_0)$ of the $2\times 2$ Green function $\bm U(t,t_0)$ as follows,
\begin{subequations}
\begin{align}
\label{trans}
&u_L(t,t_0)=u_{11}(t,t_0)-u_{12}(t,t_0)\\
&u_R(t,t_0)=u_{11}(t,t_0)+u_{12}(t,t_0).
\end{align}
\end{subequations}
Here, $u_L(t,t_0)$ corresponds to the retarded Green function of the left Majorana mode $\gamma_L$, while $u_R(t,t_0)$ corresponds to the retarded Green function of the right Majorana mode $\gamma_R$ (see the detailed derivation in Appendix B). From Eq.~(\ref{UV}), one can find that they satisfy the following integro-differential equations,
\begin{subequations}\label{umins}
\begin{align}
&\frac{d}{dt}u_L(t,t_0)=- 4 \!\! \int^t_{t_0} \!\!\! d\tau {\rm Re}[g(t,\tau)]u_L(\tau,t_0)\label{uminsa}\\
&\frac{d}{dt}u_R(t,t_0)=0 ,
\label{uminsb}
\end{align}
\end{subequations}
with initial conditions $u_L(t_0,t_0)=u_R(t_0,t_0)=1$. Equation (\ref{uminsb}) explicitly shows that $u_R(t,t_0)=1$, namely, the right Majorana mode $\gamma_R$ is decoupled from the environment so that it is decoherence-free, as we expected from Eq.~(\ref{HI3}).

Using the definition of Majorana operators Eq.~(\ref{gamma}), the master equation (\ref{ME}) can be rewritten in terms of Majorana operators $\gamma_L$ and $\gamma_R$. Notice that we do not use the path-integral approach given in Appendix A to derive directly the master equation for Majorana zero modes, because it is not obvious how Grassmann variables can be applied to Majorana operators which do not obey Fermi-Dirac statistics.
Also we find that the time-dependent dissipation and fluctuation coefficients $\lambda(t,t_0)$, $\widetilde{\lambda}(t,t_0)$ and $\Lambda(t,t_0)$ in Eq.~(\ref{ME}) are all related to each other and can be expressed in terms of $u_L(t,t_0)$ alone (see the proof given in Appendix B),
\begin{align}
\lambda(t,t_0)&=\widetilde{\lambda}(t,t_0)=-\Lambda(t,t_0)=-\frac{1}{2}\dot{u_L}(t,t_0)u_L^{-1}(t,t_0).
\label{lambda_u}
\end{align}
The identity between $\lambda(t,t_0)$ and $-\Lambda(t,t_0)$ in the master equation comes from the fact that particle-hole coupling and the particle-particle pair coupling between the zero and non-zero energy bogoliubons are given by the same coupling strength $X_{0k}$, see Eq.~(\ref{XY}), as a consequence of local perturbations. The identity between
$\lambda(t,t_0)$ and $\widetilde{\lambda}(t,t_0)$ is a result of the initial state of the TSC being the vacuum state of bogoliubons, for maintaining the TSC in superconducting states, see Eq.~(\ref{rho initial}).

Thus, using the definition Eq.~(\ref{gamma}) and the relations Eq.~(\ref{lambda_u}), the exact master equation (\ref{ME}) in terms of the Majorana operators
$\gamma_L$ and $\gamma_R$ becomes surprisingly simple,
\begin{align}
\frac{d}{dt}\rho(t)&=\lambda(t,t_0)[\gamma_L\rho(t)\gamma_L-\rho(t)].
\label{mME}
\end{align}
This master equation for Majorana zero-modes justifies the fact that gate-induced charge fluctuations
only affect on the left Majorana zero mode, as a local perturbation described by the coupling Hamiltonian
Eq.~(\ref{HI3}), and the right Majorana zero mode is decoherence-free.

To see more clearly the physical picture described by the master equation (\ref{mME}), we write the Majorana operators in the basis $\{|0\rangle,b^\dag_0|0\rangle\}$. Then $\gamma_L$ and $\gamma_R$ can be expressed in terms of Pauli matrices
\begin{align}
\label{Pauli}
& \gamma_L \rightarrow
\begin{pmatrix}
0 & -i\\
i & 0
\end{pmatrix}=\sigma_y~,~~
 \gamma_R \rightarrow
\begin{pmatrix}
0 & 1\\
1 & 0
\end{pmatrix}=\sigma_x  \notag \\
& i\gamma_L\gamma_R \rightarrow
\begin{pmatrix}
1 & 0\\
0 & -1
\end{pmatrix}=\sigma_z.
\end{align}
Namely, the Majorana operators $\gamma_L$ and $\gamma_R$ correspond to a spin operator in the zero-energy bogoliubon basis. Thus, the master equation (\ref{mME}) can be equivalently written as
\begin{align}
\frac{d}{dt}\rho(t)&=\lambda(t,t_0)[\sigma_y\rho(t)\sigma_y-\rho(t)],
\label{pure dephasing}
\end{align}
which actually describes the pure dephasing dynamics of a spin qubit. To see this explicitly, let the system be prepared in the initial state $b^\dag_0|0\rangle$. The corresponding density matrix in the basis $\{|0\rangle,b^\dag_0|0\rangle\}$ is given by $\rho(t_0)=\begin{pmatrix}0 & 0\\0 & 1\end{pmatrix}$. Then in the $\sigma_y$-representation, it becomes
\begin{align}
\rho^{(y)}(t_0)=\frac{1}{2}
\begin{pmatrix}
1 & -1\\
-1 & 1
\end{pmatrix}.
\end{align}
In this representation, the general solution of Eq.~(\ref{pure dephasing}) is
\begin{align}
\rho^{(y)}(t)=\frac{1}{2}
\begin{pmatrix}
1 & -u_L(t,t_0)\\
-u_L(t,t_0) & 1
\end{pmatrix}.
\label{rho_x}
\end{align}
The result shows that the diagonal elements of Eq.~(\ref{rho_x}) remain unchanged under the time evolution, while the off-diagonal elements will decay, as a pure dephasing. The dephasing is determined by the retarded Green function $u_L(t,t_0)$ of the left Majorana mode. In other words, Eq.~(\ref{mME}) is equivalent to the master equation of a spin-like pure dephasing model. The main difference is that here the Majorana zero mode couples to a fermionic bath, while in the usual spin dephasing model, the spin is coupled to a bosonic bath.

In conclusion, decoherence dynamics of the zero-energy bogoliubon in the model Eq.~(\ref{HT}) corresponds to the decoherence dynamics of the left Majorana mode through a local perturbation, as given by Eq.~(\ref{HI3}), and it also corresponds to a pure dephasing model in the spin basis representation. This is a very interesting result obtained in this exact master equation formalism. Also, we should emphasize that although the exact master equation we obtained here, i.e.~Eqs.~(\ref{ME}) or (\ref{mME}) has a time-convolutionless form, it is very different from Markovian master equations. The dissipation and fluctuation coefficients in our exact master equation involve deeply time convolution structures determined by the integro-differential equations (\ref{UVa}) or (\ref{uminsa}), from which non-Markovian dynamics are fully taken into account through the integral memory kernel in Eq.~(\ref{UVa}) or (\ref{uminsa}). While, Markovian master equations, such as the usual Lindblad master equation derived from semigroup maps [\onlinecite{Lindblad1976}], do not have such a convolution structure to capture non-Markovian dynamics.

\section{Non-Markovian Decoherence Dynamics of Majorana zero modes}
%

The master equation formalism obtained in the last section allows us to describe in general the exact non-Markovian decoherence dynamics of Majorana zero modes under the gate-induced charge fluctuations. The time nonlocal dissipation and fluctuation coefficients in the master equation are fully determined by the nonequilibrium retarded Green function $u_L(t,t_0)$ in this model, see Eq.~(\ref{lambda_u}), which depicts all possible non-Markovian memory dynamics.
Explicitly, as shown in our previous work [\onlinecite{PRL2012}], the general non-Markovian solution of $u_L(t,t_0)$, solved exactly from the integro-differential equation (\ref{uminsa}), consists of a nonexponential decay and a summation of dissipationless oscillations,
\begin{align}
u_L(t,t_0)=&\sum_b \frac{1}{1-\Sigma'(\omega_b)}e^{-i\omega_b (t-t_0)}\notag\\
&+\!\! \int^\infty_{-\infty}\!\!\!\!\! \!\! d\omega \frac{{\cal J}(\omega)e^{-i\omega (t-t_0)}}{[\omega-\delta\omega]^2+\pi^2{\cal J}^2(\omega)},
\label{uplus}
\end{align}
where
\begin{align}
{\cal J}(\omega)=2[J(\omega) + J(-\omega)]   \label{esd}
\end{align}
is the effective spectral density of the charge fluctuations on the left Majorana zero mode.

The dissipationless oscillations, the first term in the above solution, are contributed from localized bound states, resulting from band gaps or zero-points in the environment spectral density as well as the system-environment coupling strength. Such dissipationless oscillations provide a long-time non-Markovian memory process, namely they carry a partial  information of the initial state forever [\onlinecite{PRL2012,PRA2015}]. The localized bound state energy $\omega_b$ is determined by the following pole condition,
\begin{align}
\omega_b-\delta\omega_b=0~{\rm with}~{\cal J}(\omega_b)=0,
\label{polecondition}
\end{align}
where
\begin{align}
\delta\omega=\mathcal{P}\int^\infty_{-\infty} d\omega'\frac{{\cal J}(\omega)}{\omega-\omega'}  \label{enegs}
\end{align}
and $\mathcal{P}$ denotes a principal-value integral. The energy shift $\delta \omega$ is the real part of the self-energy correction to the left Majorana zero mode, induced by charge fluctuations. This self-energy correction, i.e.~the Fourier transformation of the integral kernel in Eq.~(\ref{uminsa}) is given by
\begin{align}
\Sigma(\omega\pm i0^+)&= \! \int^\infty_{-\infty} \!\!\!\! d\omega'\frac{{\cal J}(\omega)}{\omega-\omega'\pm i0^+}
=\delta\omega\mp i \pi {\cal J}(\omega) .
\label{selfenergy}
\end{align}
The prefactor $[1-\Sigma'(\omega_b)]^{-1}$ in the solution (\ref{uplus}) is the residue at the pole $\omega_b$.

The nonexponential decay term, the second term in Eq.~(\ref{uplus}), comes from the discontinuity in the imaginary part of the environmental-induced self-energy correction to the system [see Eq.~(\ref{selfenergy})]. This nonexponential decay term can result in dissipation coefficient $\lambda(t,t_0)$ oscillating between positive and negative values in short times, as a short-time non-Markovian memory effect [\onlinecite{PRL2012,PRA2015}]. Furthermore, in the weak charge fluctuation regime, $\lambda(t,t_0)$ becomes always positive, and even could be a constant as a Markov limit. This
general picture of decoherence dynamics can be further specified by calculating the dissipation (decay) coefficient $\lambda(t,t_0)$ from the solution Eq.~(\ref{uplus}) via Eq.~(\ref{lambda_u}).

In the following subsections, we will present in detail the analytical as well as numerical solutions of non-Markovian decoherence dynamics to the left Majorana mode, induced by the controlling gate at zero and finite temperatures, respectively, and then
provide a comparison with the Markovian dynamics obtained from the second-order perturbation in Ref.~[\onlinecite{Daniel2012}].
\subsection{Non-Markovian decoherence dynamics at zero temperature}
\begin{figure}
\centerline{\scalebox{0.25}{\includegraphics{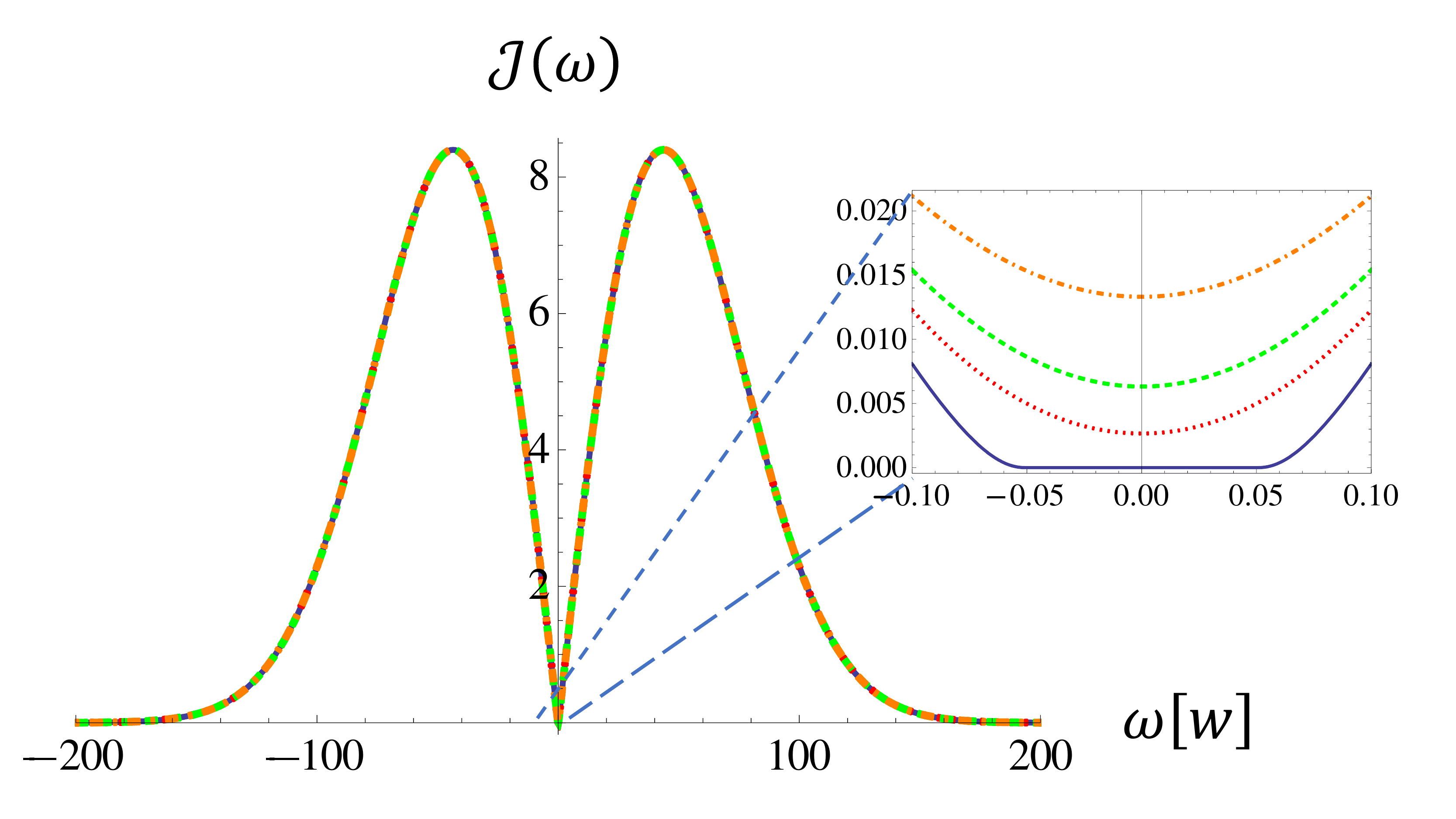}}}
\caption{(color online) The effective spectral density ${\cal J}(\omega)=2[J(\omega)+J(-\omega)]$ with charge fluctuation strength $\frac{\pi}{2}B_G\eta^2=1.0$ at different gate temperatures: $k_BT=0$ (solid), $0.025w$ (dotted), $0.035w$ (dashed), $0.05w$ (dotted-dashed). The Fermi energy $E_F=100w$ and the cut-off frequency $\omega_c=5w$ and the gap parameter $\Delta=0.05w$. Note that ${\cal J}(\omega)=0$ for $|\omega|<\Delta$ at $T=0$. }\label{spectrum}
\end{figure}

At zero temperature ($T=0$), the gate spectral density  $J_G(\omega)$ is zero for $\omega<0$. This leads to the effective spectral density ${\cal J}(\omega)=0$ for $|\omega|<\Delta$ [see Eqs.~(\ref{JT}) and (\ref{esd}) and Fig.~\ref{spectrum}]. On the other hand, $\delta \omega$ is an odd function of $\omega$, see Eq.~(\ref{enegs}). As a result, the pole condition Eq.~(\ref{polecondition})
is only satisfied by $\omega_b=0$. In other words, at zero temperature, the general solution Eq.~(\ref{uplus}) is reduced to
\begin{align}
u_L(t, t_0)& =\frac{1}{1-\Sigma'(0)}
+\! \int^\infty_{\Delta} \!\!\!\!\!\!  d\omega \frac{2{\cal J}(\omega)\cos[\omega (t-t_0)]}{[\omega-\delta\omega]^2+\pi^2{\cal J}^2(\omega)},
\label{u_L}
\end{align}
which contains a zero-energy localized bound state, the first term that is dissipationless as a long-time non-Markovian memory effect, and a decay dynamics given by the second term as a level broadening effect. The latter precisely describes the decoherence (amplitude damping) of the left Majorana mode.
Note that
$u_L(t,t_0)$ is a real function, and as we expected in the last section that the renormalized zero-energy bogoliubon remains to have zero energy.

In fact, the zero-energy localized bound state in Eq.~(\ref{u_L}) represents only a part of the original Majorana zero mode. To see this clearly, we may write Eq.~(\ref{u_L}) by $u_L(t,t_0)= \int \! d\omega {\cal D}(\omega)e^{-i\omega (t-t_0)}$, where
\begin{align}
{\cal D}(\omega) = \frac{1}{1\!-\!\Sigma'(\omega)}\delta(\omega) \!+ \!\frac{{\cal J}(\omega)}{[\omega\!-\!\delta\omega]^2\!+\!\pi^2{\cal J}^2(\omega)}.
\end{align}
It shows how charge fluctuations induce level broadening to the left Majorana zero mode.
The spectrum ${\cal D}(\omega)$ of the renormalized left Majorana mode still
has a peak at zero-energy, but the amplitude of this zero-mode becomes less than one, $[1-\Sigma'(0)]^{-1}<1$, namely the left Majorana zero mode does decay (amplitude damping) even at $T=0$. The decay occurs because charge fluctuations induce the level broadening, the continuous part of the spectrum represented by the second term in the above solution.
The decay rate  of the left Majorana zero mode, which is just the dissipation coefficient $\lambda(t,t_0)$ in the master equation,
can then be calculated directly from Eq.~(\ref{lambda_u}) which can also be expressed as
\begin{align}
\lambda(t,t_0)
&=2 \!\! \int^t_{t_0} \!\!\! d\tau{\rm Re}[g(t,\tau)]u_L(\tau,t_0)u^{-1}_L(t,t_0) .  \label{lambda}
\end{align}

\begin{figure}
\centerline{\scalebox{0.25}{\includegraphics{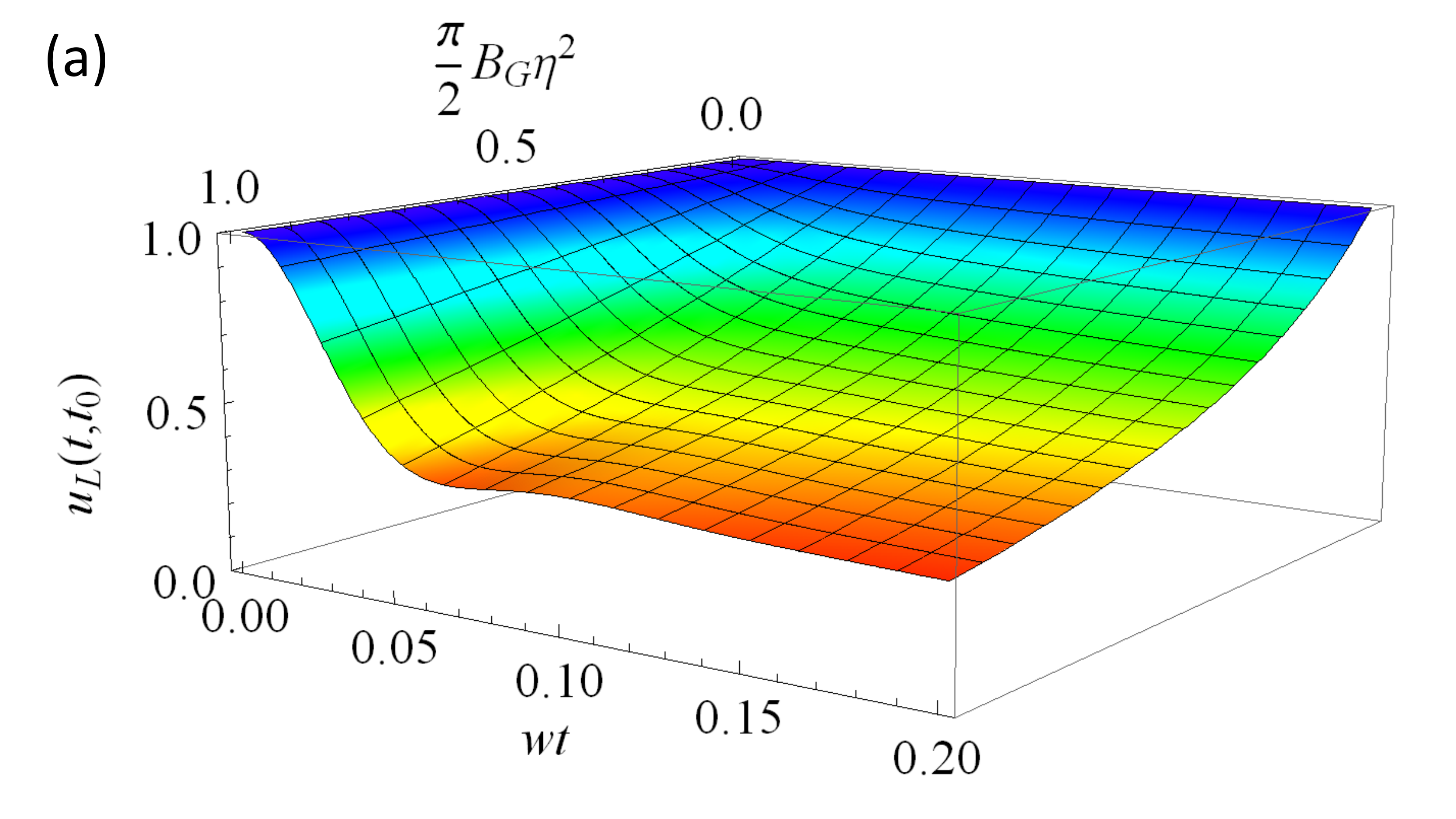}}}
\centerline{\scalebox{0.26}{\includegraphics{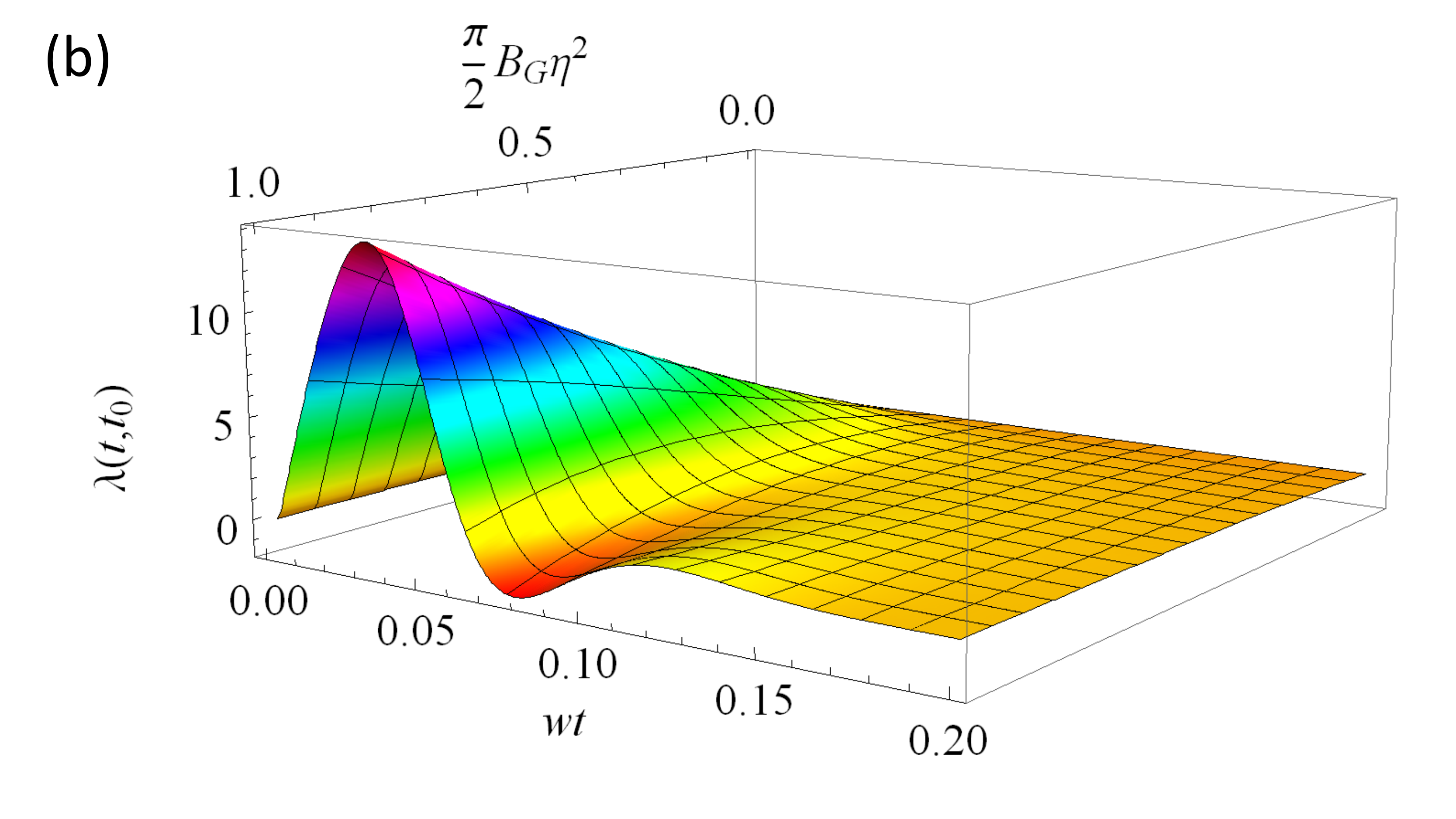}}}
\caption{(color online) The retarded Green function $u_L(t,t_0)$ and the decay rate $\lambda(t,t_0)$ which characterize
the left Majorana zero mode decoherence, are plotted as a function of the dimensionless time $wt$ and the charge fluctuation strength $\frac{\pi}{2}B_G\eta^2$ at $T=0$. The gap parameter $\Delta=0.05w$ here. } \label{unoppzt}
\end{figure}

To show explicitly the decoherence dynamics of the left Majorana mode, we plot in Fig.~\ref{unoppzt} the time-dependence
of $u_L(t,t_0)$ and the time-dependent decay rate $\lambda(t,t_0)$
as a function of charge fluctuation strength $\frac{\pi}{2}B_G\eta^2$ at $T=0$.
We take the parameters given in Ref.~[\onlinecite{Daniel2012}], which are close to a reliable physical situation,
as pointed out in Ref.~[\onlinecite{Daniel2012}].
It shows in Fig.~\ref{unoppzt} that when the charge fluctuation is very weak, namely the factor $\frac{\pi}{2}B_G\eta^2$ is very small, e.g.~$\frac{\pi}{2}B_G\eta^2=0.01$, the decoherence of the left Majorana zero mode is actually very small and almost negligible.
Correspondingly, the decay rate $\lambda(t,t_0)$ is almost close to zero.
Increasing the strength of charge fluctuations will induce decoherence to the left Majorana zero mode, namely the value of $u_L(t,t_0)$
will decay away quickly from unity. The decay rate $\lambda(t,t_0)$ also increases accordingly.
When the charge fluctuations become stronger, e.g.~$\frac{\pi}{2}B_G\eta^2=1$, the damping of the left Majorana zero mode
becomes significant, as shown in Fig.~\ref{unoppzt}. In other words, Majorana zero modes are not so robust against
decoherence as one was expected, even at zero temperature.

More interestingly, Fig.~\ref{unoppzt}b shows that for strong charge fluctuations
$\frac{\pi}{2}B_G\eta^2 \gtrsim 0.6 $, the decay rate has an oscillation between positive and negative
values in the short-time regime. The positive values of $\lambda(t,t_0)$ describe
the decoherence of the left Majorana mode, namely the zero-energy bogoliubon is excited into finite-energy bogoliubon states,
while the negative values of $\lambda(t,t_0)$ describe the inverse dissipation
from the environment back into the system (a feedback effect). Hence, an oscillating  $\lambda(t,t_0)$ between positive and negative values depicts nonequilibrium memory processes, which is defined as a typical short-time non-Markovian effect in our previous works
[\onlinecite{PRA2015}]. On the other hand, Fig.~\ref{unoppzt}a shows that $u_L(t,t_0)$ does not decay to zero
even in the strong coupling regime.
This implies that the left Majorana zero mode is only partially decoherent. This partial decoherence is due to the existence of the
localized bound state at zero energy, the first term in Eq.~(\ref{u_L}) which is dissipationless, a long-time non-Markovian memory effect [\onlinecite{PRL2012}] that partially protects the left Majorana zero mode from a complete decoherence.

Also, from the Green function $u_L(t,t_0)$, one can calculate the unoccupied probability of the zero-energy bogoliubon
\begin{align}
P(t,t_0)=1-\langle b^{\dag}_0(t)b_0(t)\rangle=\frac{1}{2}[1-u_L(t,t_0)]   \label{unoccupied}
\end{align}
for the system being initially in the state $b^\dag_0|0\rangle$. Equation (\ref{unoccupied}) makes an explicit connection about the decoherence between the zero-energy bogoliubon and the left Majorana zero mode. Because a Majorana zero mode represents a half of one electron, and only the left Majorana zero mode is disturbed under local perturbations (the right Majorana zero mode is decoherence-free), the unoccupied probability can range from $0$ to $0.5$. This is justified precisely in Eq.~(\ref{unoccupied}) by the fact that $u_L(t,t_0)$ varies from $1$ (no decoherence)
to $0$ (complete decoherence). Thus, $u_L(t,t_0) = 1$ means no decoherence, correspondingly $P(t,t_0)=0$; while  $0<u_L(t,t_0)<1 $ corresponds to partial decoherence so that $0< P(t,t_0)<0.5$. If $u_L(t,t_0)=0$, it implies a complete decoherence for the left Majorana zero mode, and $P(t,t_0)=0.5$. Therefore, decoherence of the left Majorana zero mode fully
characterizes the decoherence of the zero-energy bogoliubon, which justifies the consistency of the master equations between (\ref{ME}) and (\ref{mME}).


\subsection{Non-Markovian decoherence dynamics at finite temperature}
Note that the gate temperature $T$ in Eq.~(\ref{JG}) is not necessary to be the temperature of the TSC.
However, because the gate is applied to the TSC, to maintain the TSC in a superconducting state, the gate
temperature should be not too high. In general, we may assume that the gate temperature $T$ is lower than
the critical temperature $T_c$ of the TSC.
Then, at a finite temperature ($T < T_c$), it shows that although the effective spectral density ${\cal J}(\omega)\neq 0$ over the whole frequency
range, there is a sharp dip near the zero frequency, corresponding to ${\cal J}(0)$ which is not zero but very small, see the inset in Fig.~\ref{spectrum}. In other words, the localized bound state at zero energy becomes a resonant state with a very small linewidth $\simeq \pi{\cal J}(0)$, approximately given by
$\omega_b \simeq 0 + i \pi{\cal J}(0)$ in Eq.~(\ref{u_L}).
This indicates that the decoherence of the left Majorana zero mode at $T\neq 0$ is almost the same as
the decoherence at $T=0$, dominated by the second term in Eq.~(\ref{u_L}), plus an additional weak decay caused by
the small value ${\cal J}(\omega)$ in the regime $|\omega| < \Delta$ for a finite temperature effect.

\begin{figure}
\centerline{\scalebox{0.45}{\includegraphics{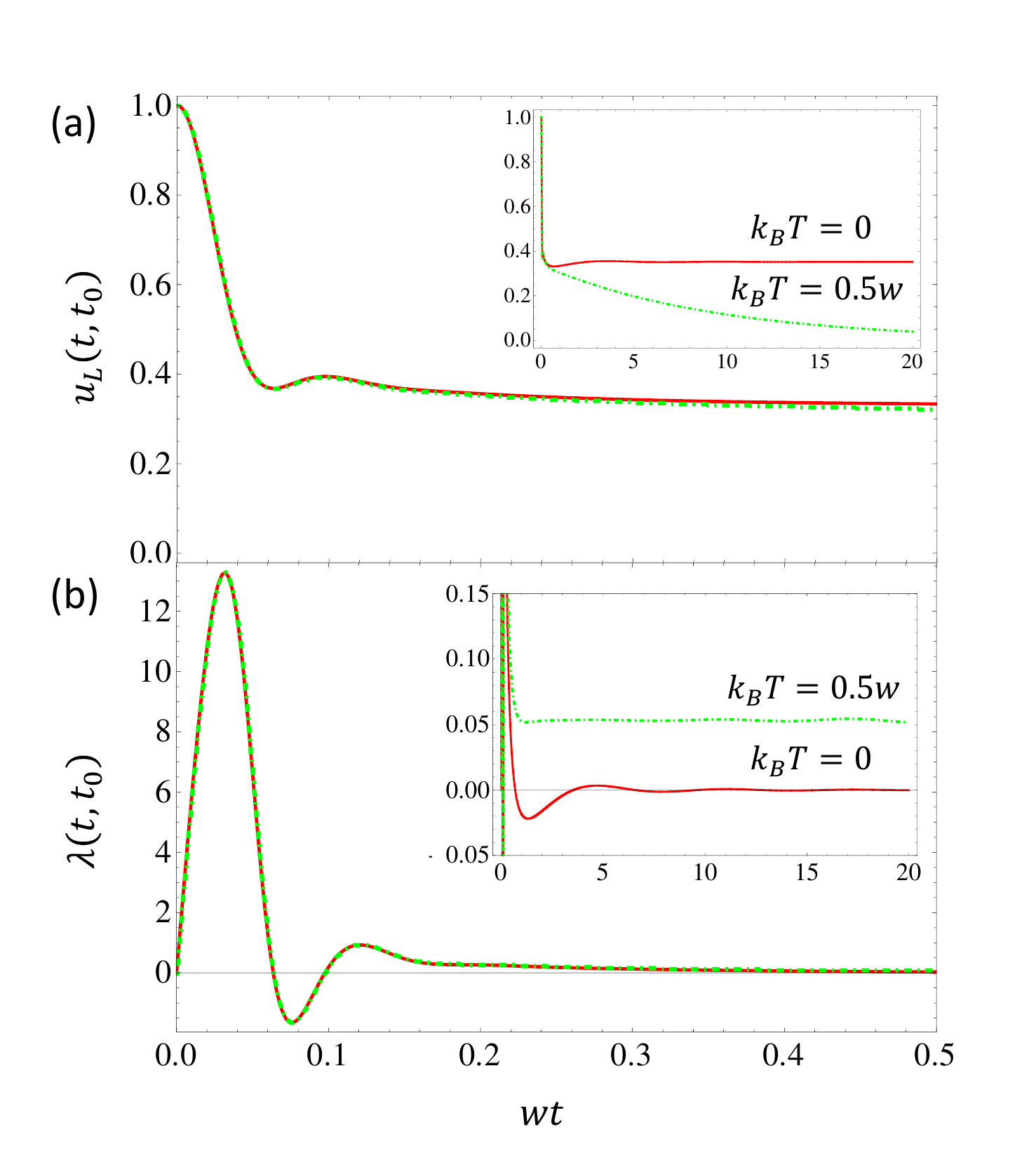}}}
\caption{(color online) (a) The retarded Green function $u_L(t,t_0)$ and (b) the decay rate $\lambda(t,t_0)$
are plotted as a function of the time $wt$ at zero (red-solid) and finite (green-dotted) temperatures.
Here the charge fluctuation strength $\frac{\pi}{2}B_G\eta^2=1.0 $ and the gap parameter $\Delta=w$.}\label{lambdaft}
\end{figure}

The 3D plots of the time-dependence of the Green function $u_L(t,t_0)$ and the time-dependent decay rate $\lambda(t,t_0)$
at $T\neq 0$, as a function of charge fluctuation strength $\frac{\pi}{2}B_G\eta^2$, are very similar to
the case presented in Fig.~\ref{unoppzt} for $T=0$, except for a little difference in the long time region.
We compare this decoherence difference by plotting $u_L(t,t_0)$ and $\lambda(t,t_0)$ in Fig.~\ref{lambdaft} for the
charge fluctuation strength $\frac{\pi}{2}B_G\eta^2=1.0$ with the ideal case of the gap parameter being the
same as the hopping amplitude $w$: $\Delta=w$.
The results show that both the time evolution of the Green function $u_L(t,t_0)$ and the time-dependent decay
rate $\lambda(t,t_0)$ are qualitatively the same for zero (red-solid) and finite (green-dotted) temperatures. The only difference is that  there
is a finite steady-state value of $\lambda(t,t_0)$ at $T\neq 0$ for large $t$, see the inset in Fig~\ref{lambdaft}b.
This small value of $\lambda(t,t_0)$ at large $t$ comes from the nonzero ${\cal J}(\omega)$ in the sharp dip of
the spectral density near zero frequency, see Fig.~\ref{spectrum}. It makes the left Majorana mode
completely decoherence in a long time limit, see Fig~\ref{lambdaft}a.
In other words, the zero-energy bogoliubon at
$T\neq 0$ will be totally excited to non-zero bogoliubon modes eventually,
even though this gate temperature is lower than the superconducting critical temperature.
We also find that such a finite steady-state value of $\lambda(t,t_0)$ increases when the temperature increases,
which implies a faster decoherence of the Majorana zero mode at higher temperature.

\subsection{Comparison between the exact non-Markovian dynamics and the second-order perturbation solution (Markovian approximation)}
As we have pointed out in the introduction, the decoherence dynamics of the zero-energy bogoliubon with the model Hamiltonian Eq.~(\ref{HT}) was studied by Schmidt {\it et al.}~[\onlinecite{Daniel2012}], using Fermi's golden rule. In this subsection, we show that their results can be obtained directly by taking the second-order perturbation approximation from our exact formalism. Explicitly, integrating over the time in both sides of Eq.~(\ref{uminsa}), and then taking a perturbation expansion with respect to the charge fluctuation strength $\frac{\pi}{2}B_G\eta^2   \sim g(t,\tau)$, one has
\begin{align}
u_L(t,t_0) = 1& -4\!\! \int^t_{t_0}\!\!\!  d\tau_1 \!\! \int^{\tau_1}_{t_0} \!\!\!\!\! d\tau_2 {\rm Re}[g(\tau_1,\tau_2)] u_L(\tau_2, t_0) \notag\\
= 1& - 2 \!\! \int^t_{t_0} \!\!\! d\tau_1 \!\! \int^t_{t_0} \!\!\!  d\tau_2  g(\tau_1,\tau_2)  \notag \\
& + 2^2 \!\! \int^t_{t_0} \!\!\! d\tau_1 \!\! \int^t_{t_0} \!\!\! d\tau_2 \!\! \int^{\tau_2}_{t_0}\!\!\!\! d\tau_3 \!\! \int^{\tau_2}_{t_0} \!\!\!\! d\tau_4 g(\tau_1,\tau_2) g(\tau_3,\tau_4) \notag \\
& + \cdots.
\label{uL_perturb}
\end{align}
Make use of the second-order approximation from the above perturbation expansion, the unoccupied probability of the zero-energy bogoliubon Eq.~(\ref{unoccupied}) is reduced to
\begin{align}
P&(t,t_0) \approx \!\! \int^t_{t_0}\!\!\!  d\tau_1 \!\! \int^t_{t_0} \!\!\! d\tau_2g(\tau_1,\tau_2)\notag\\
&=\eta^2\sum_{k\neq0}X^2_{0k} \!\! \int^t_{t_0} \!\!\!  d\tau_1 \!\! \int^t_{t_0} \!\!\!  d\tau_2 \langle\delta Q(\tau_1)\delta Q(\tau_2)\rangle_G e^{-iE_k(\tau_1-\tau_2)} .
\end{align}
This is the result given by Eq.~(18) in Ref.~[\onlinecite{Daniel2012}].

Also, making the second-order approximation to the decay coefficient of Eq.~(\ref{lambda}), we have
\begin{align}
\lambda(t,t_0)&\approx 2 \! \int^t_{t_0} \!\!\! d\tau{\rm Re}[g(t,\tau)]\notag\\
&=2 \! \int^t_{t_0}  \!\!\! d\tau \!\! \int^{\infty}_{-\infty}  \!\!\!\!\! d\omega J(\omega){\rm cos}[\omega(t-\tau)].
\end{align}
If we take further the long-time limit ($t \! \rightarrow \! \infty$), the second-order approximation of the decay coefficient $\lambda(t,t_0)$ becomes
\begin{align}
\lambda&(t \! \rightarrow \! \infty, t_0)\approx 2\pi J(0)=2\pi\eta^2\sum_{k\neq0}X^2_{0k}J_G(-E_k)\notag\\
&=2\pi B_G\eta^2\sum_{k\neq0}X^2_{0k}\exp \!\Big(\!\!-\!\!\frac{E_k^2}{8E_F \omega_c}\Big)\frac{E_k}{e^{E_k/kT}-1} .
\label{drlt}
\end{align}
This produces the decay constant (inverse of the life-time) given by Eq.~(22) in Ref.~[\onlinecite{Daniel2012}].
As we have shown in our previous works [\onlinecite{Xiong2010,PRA2015}], the Markovian dynamics of open systems can be obtained from our exact master equation, by taking the second-order perturbation with respect to the system-environment coupling, the resulting master equation is the Born-Markov master equation  [\onlinecite{Xiong2010}].
Therefore, the above second-order perturbation solution can only describe Markovian dynamics, the long time limit corresponds to Markov limit.

\begin{figure}
\centerline{\scalebox{0.45}{\includegraphics{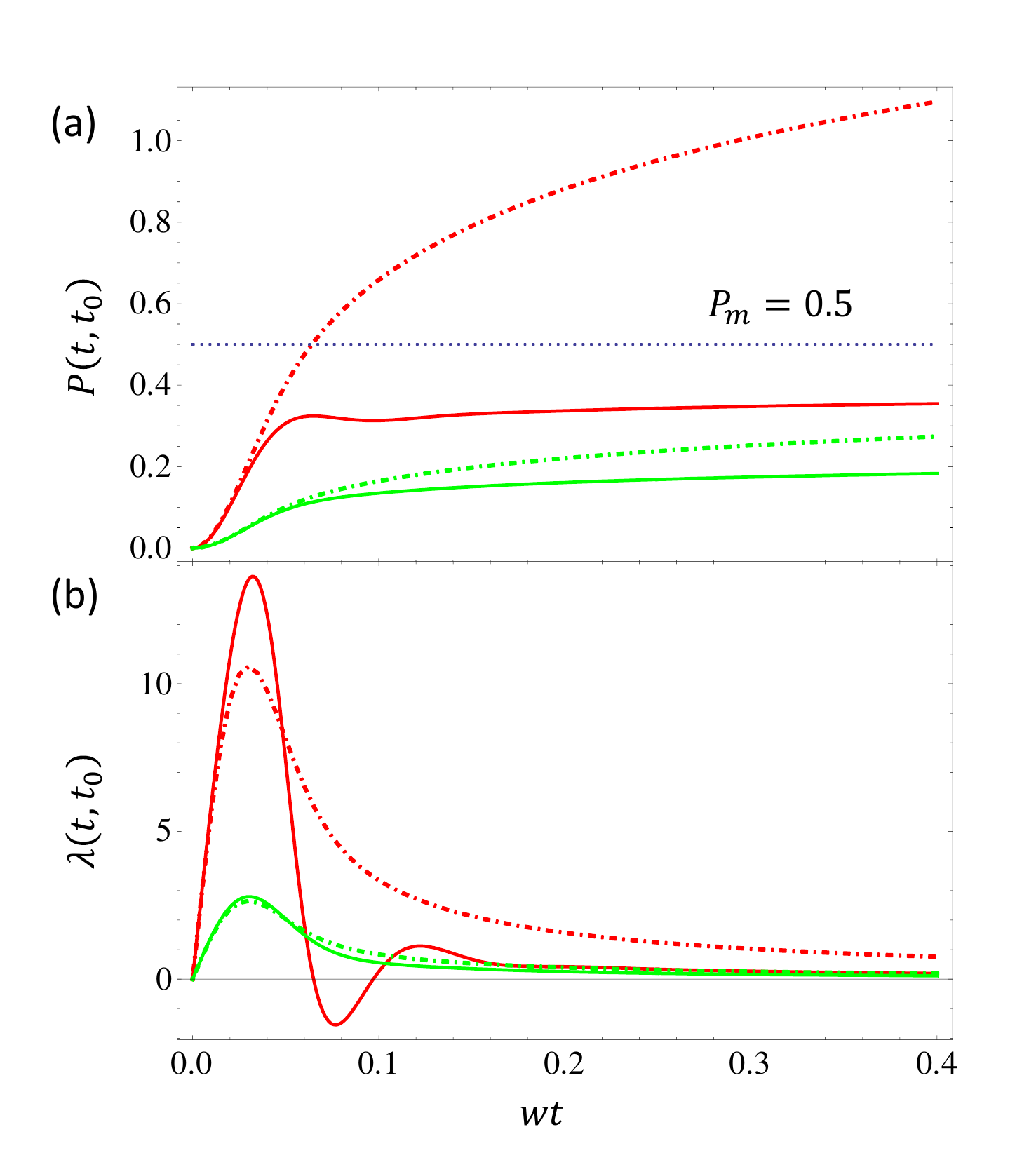}}}
\caption{(color online) The comparison of the exact solution (solid lines) with the second-order approximation (dotted-dashed lines) for the time-evolution of the unoccupied probability $P(t,t_0)$ and the decay rate $\lambda(t,t_0)$ at $T=0$, with charge fluctuation strength $\frac{\pi}{2}B_G\eta^2=0.25$ (Green) and $\frac{\pi}{2}B_G\eta^2=1$ (Red), the gap parameter $\Delta=0.05w$. For exact solutions (solid lines), in plot (a) a long-time non-Markovian memory effect shows up for both the strong and weak charge fluctuations, corresponding to the steady values ($<0.5$) resulting from the dissipationless localized bound state. In plot (b), a short-time non-Markovian dynamics shows up for strong charge fluctuation strength, corresponding to $\lambda(t,t_0)$ oscillating between positive and negative values, as information flowing back memory effect. The second-order perturbation (dotted-dashed lines), as a Markovian approximation, cannot produce such effects. } \label{Gam2D}
\end{figure}

In Fig.~\ref{Gam2D} we make a comparison between the exact solution and the second-order perturbation result at $T=0$.
Figure \ref{Gam2D}a is the time evolution of the unoccupied probability of the zero-energy bogoliubon.
It shows that for both the weak and strong charge fluctuations, the exact solution of $P(t,t_0)$
grows up in a very short-time evolution region $wt < 0.08$, i.e.~the zero-energy bogoliubon is partially excited, resulting from the second term in Eq.~(\ref{u_L}) as a decoherence  of the left Majorana zero mode. After that, $P(t,t_0)$ approaches to a steady
value contributed from the localized bound state, the first term of Eq.~(\ref{u_L}) that protects the left Majorana zero mode
from any further decoherence, as a general long-time non-Markovian memory effect that we have discussed in Sec.~IIIA.
It also shows that $P(t,t_0)$ is always less than $0.5$ in the exact solution, namely the left Majorana zero mode is never
fully decohered away at $T=0$, due to the existence of the zero-energy localized bound state.
However, the results from the second-order perturbation shows that for both weak and strong
charge fluctuations, $P(t,t_0)$ behaviors similar to the exact solution only in the very short-time evolution
range, $wt < 0.08$. After that, $P(t,t_0)$ keeps growing up linearly in time and then quickly exceeds
the physically allowed maximum value $P_m=0.5$.

Figure \ref{Gam2D}b is the time evolution of the decay rate $\lambda(t,t_0)$. The results show that when the charge fluctuation strength is weak, the second-order approximation is close to the exact result. When the charge fluctuation strength becomes strong, the exact decay rate $\lambda(t,t_0)$ shows negative values at transient regime, indicating a short-time non-Markovian memory effect, as we have also discussed in Sec. IIIA. However, the second-order approximation of $\lambda(t,t_0)$ always gives a positive value, which corresponds to a Markovian process. This justified the fact that the second-order approximation ignores all possible non-Markovian memory effect in the decoherent dynamics [\onlinecite{PRA2015}].

It may be also worth pointing out that the unoccupied probability $P(t,t_0)$ is the transition probability that the zero-energy bogoliubon is
excited into finite-energy bogoliubon states. As it is well-known, the transition probability is linearly proportional to $t$
in the second-order perturbation theory, from which one can determine the decay rate (the transition rate)
as a result of Fermi's golden rule [\onlinecite{Sakurai1994}]. However, here the unoccupied probability $P(t,t_0)$
has a maximum value, $P_m=0.5$, which corresponds to the left Majorana zero mode be completely decohered.
Thus, after $P(t,t_0)$ reaches $0.5$, no further physical processes can happen. Figure \ref{Gam2D}a
shows that as time evolves, the unoccupied probability in the second-order perturbation theory will exceed
the physically allowed maximum value of $0.5$ in large $t$.
On the other hand, the decay rate determined from Fermi's
golden rule [\onlinecite{Daniel2012}] is indeed the long-time limit of the decay rate in the second-order
perturbation theory, see Eq.~(\ref{drlt}). In other words, the decay rate calculated from Fermi's golden rule
is the value at the tails of these curves in Fig.~\ref{Gam2D}b. It does not characterize the true decays
happened in the short-time region (corresponds to $ 0< wt \lesssim 0.15$ in Fig.~\ref{Gam2D}b) which is actually the dominant decoherence time region, in particular, for non-Markovian dynamics. Only for the very weak charge fluctuations, the decay rate obtained from Fermi's golden rule is close to the time-dependent (both the exact and the second order perturbation) solution, see Fig.~\ref{unoppzt}b and Fig.~\ref{Gam2D}b.

\section{Decoherence dynamics of two-time correlation functions}
In this section, we shall study the real-time correlations of Majorana zero modes.
As we can see from Eq.~(\ref{Pauli}), the two-time correlation functions $\langle\gamma_L(t+\tau)\gamma_R(t)\rangle$, $\langle\gamma_R(t+\tau)\gamma_L(t)\rangle$, $\langle\gamma_L(t+\tau)\gamma_L(t)\rangle$ and $\langle\gamma_R(t+\tau)\gamma_R(t)\rangle$ correspond to $\sigma_x$ and $\sigma_y$ operations at different times. In practice, a topological qubit should be made by at least four Majorana zero modes so that not only the qubit flip but also the qubit coherence can be fully manipulated [\onlinecite{Nature2011,Goldstein2011}], the corresponding decoherence dynamics remains for our next investigation. Here
we will focus on the decoherence dynamics contained in the above listed two-time correlation functions, under the local perturbations given by Eq.~(\ref{HT}). The physical picture of the two-time correlation functions is given as follows: when the system is initially prepared in a given state, it starts to evolve until an operation is performed at the operation time $t$. After then, the system continues to evolve from time $t$ to time $t+\tau$, and another operation is performed at time $t+\tau$. Here $\tau$ is usually called the delay time. These correlation functions also provide information of the memory persistence of Majorana zero modes between times $t$ and $t+\tau$ [\onlinecite{Goldstein2011}, \onlinecite{PRA2015}]. Furthermore, these two-time correlation functions are related to the nonequilibrium Green functions as follows (see Appendix B),
\begin{subequations}
\begin{align}
\langle\gamma_R(t+\tau)\gamma_R(t)\rangle=&\langle\gamma_R(t_0)\gamma_R(t_0)\rangle=1\label{CRR} \\
\langle\gamma_R(t+\tau)\gamma_L(t)\rangle=&u_L(t,t_0)\langle\gamma_R(t_0)\gamma_L(t_0)\rangle\notag \\
=&\langle\gamma_R(t)\gamma_L(t)\rangle\label{CRL}\\
\langle\gamma_L(t+\tau)\gamma_R(t)\rangle=&u_L(t+\tau,t_0)\langle\gamma_L(t_0)\gamma_R(t_0)\rangle\label{CLR} \notag \\
=&\frac{u_L(t+\tau,t_0)}{u_L(t,t_0)}\langle\gamma_L(t)\gamma_R(t)\rangle \\
\langle\gamma_L(t+\tau)\gamma_L(t)\rangle=&u_L(t+\tau,t_0)u_L(t,t_0)+v_L(t,t+\tau),\label{CLL}
\end{align}
\end{subequations}
where $u_L(t,t_0)$ is the retarded Green function discussed in the last section, and $v_L(\tau,t)$ is the correlation Green function that
will be given later, see Eq.~(\ref{cgf}).

As expected, $\langle\gamma_R(t+\tau)\gamma_R(t)\rangle=1$ because $\gamma_R$ does not couple to the environment, while $\langle\gamma_R(t+\tau)\gamma_L(t)\rangle$ describes one kind of the correlations between the left and right Majorana modes. Also, because of the decoherence-free property of the right Majorana zero mode, the decoherence of $\langle\gamma_R(t+\tau)\gamma_L(t)\rangle$ is determined only by the left Majorana mode at time $t$, which is fully determined  by the retarded Green function $u_L(t,t_0)$ that characterizes the decoherence dynamics of the left Majorana mode alone. It is independent of the delay time $\tau$ operated by the right Majorana mode.
On the other hand, it is interesting to see that another correlation between the left and right Majorana modes, $\langle\gamma_L(t+\tau)\gamma_R(t)\rangle$, which exchanges the time ordering of the correlation $\langle\gamma_R(t+\tau)\gamma_L(t)\rangle$, shows different decoherence dynamics, see Eqs~(\ref{CRL}) and (\ref{CLR}). The decoherence dynamics of the two-time correlation (\ref{CLR}) depends on both the delay time $\tau$ and the operation time $t$. We can introduce a correlation difference factor $C_{LR}(\tau,t)$ to characterize the different decoherence behavior between $\langle\gamma_R(t+\tau)\gamma_L(t)\rangle$ and $\langle\gamma_L(t+\tau)\gamma_R(t)\rangle$,
\begin{align}
C_{LR}(\tau,t) \equiv \frac{u_L(t+\tau,t_0)}{u_L(t,t_0)} .
\end{align}
Although the decoherence dynamics is fully determined by the charge fluctuations of the gate on the left Majorana mode, $C_{LR}(\tau,t)$ presents the different time dependence of decoherence.
It reveals a fact that the decoherence induced by local perturbations may affect globally (or topologically) Majorana operations, as long as $C_{LR}(\tau,t) \neq 1$.  This global decoherence property for Majorana operations is actually also unusual, because one usually believes that topological quantum computing with Majorana zero modes is robust against decoherence under local perturbations. The above solution seems to show that this common belief may not be true in practice.

\begin{figure}
\centerline{\scalebox{0.25}{\includegraphics{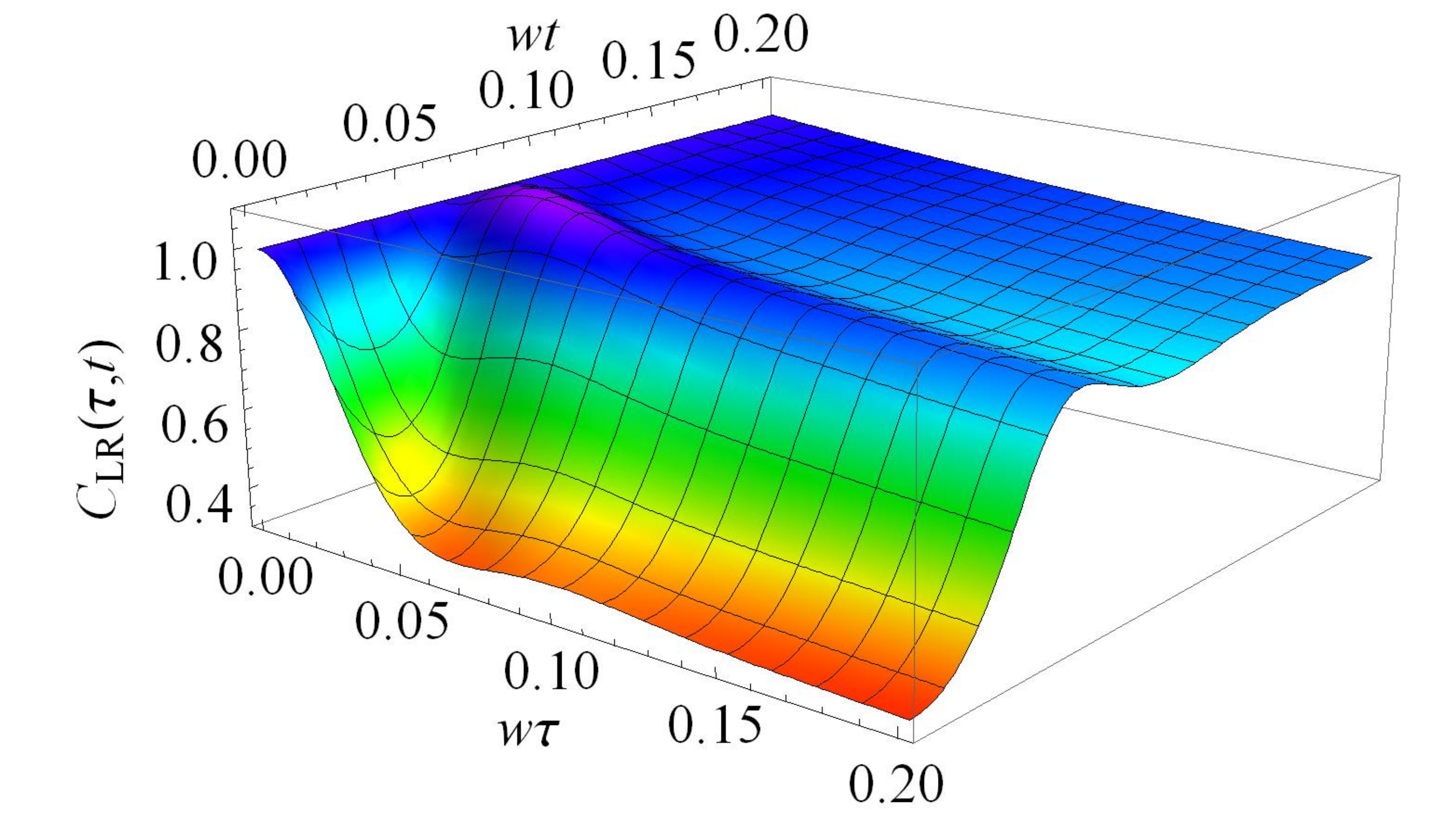}}}
\caption{(color online) The correlation different factor $C_{LR}(\tau,t)$ as a function of the operation time $wt$ and
the delay time $w\tau$
with the charge fluctuation strength $\frac{\pi}{2}B_G\eta^2=1$  at $T=0$. The gap parameter
$\Delta=0.05w$. }\label{uLL0kT}
\end{figure}

To see more explicitly, we plot in Fig.~\ref{uLL0kT} the correlation difference factor $C_{LR}(\tau,t)$ as a function of two times, the operation time $t$ and the  delay time $\tau$ at $T=0$.
As one can see from Fig.~\ref{uLL0kT}, the delay time dependence of the correlation difference is significant in a short-time regime from the beginning, originated from the short-time non-Markovian dynamics shown in Fig.~\ref{unoppzt}. After then, $C_{LR}(\tau,t)$ becomes almost $\tau$-independence. This steady-state $\tau$-independent behavior comes from the dissipationless property of the zero-energy localized bound state of the left Majorana mode at $T=0$, see Eq.~(\ref{u_L}).
On the other hand, the operation time $t$-dependence shows the transient decoherence dynamics of the left Majorana mode. When the operation time reaches the steady-state limit, the correlation difference factor $C_{LR}(\tau,t) \simeq 1$,
namely the decoherence pattern of the two-time correlation functions $\langle\gamma_R(t+\tau)\gamma_L(t)\rangle$
and $\langle\gamma_L(t+\tau)\gamma_R(t)\rangle$ becomes almost identical in the steady-state limit.
It shows that even at zero temperature, charge fluctuations can also affect globally Majorana zero modes in the transient
decoherence dynamic regime, but this is already very crucial for the reliability of quantum information processing using
Majorana zero modes. The advantage of topologically protecting Majorana zero mode operations from the local perturbation
that one expected seems to be not available.

The last two-time correlation function $\langle\gamma_L(t+\tau)\gamma_L(t)\rangle$ of Eq.~(\ref{CLL})
characterizes the decoherence dynamics of the correlation function of the left Majorana mode itself at two different times. On the one hand, this correlation measures the non-Markovian memory effect [\onlinecite{PRA2015}], and on the other hand, describes equivalently the correlation $\langle\sigma_y(t+\tau)\sigma_y(t)\rangle$. From Eq.~(\ref{CLL}), it shows that this correlation function is determined not only by the retarded Green function $u_L(t+\tau,t_0)u_L(t,t_0)$ but also the pairing-induced vacuum quantum fluctuation $v_L(t,t+\tau)$ on the left Majorana mode,
\begin{align}
v_L&(t,t+\tau)\notag\\
&=\!\! \int^t_{t_0}\!\!\!\! d\tau_1\!\! \int^{t+\tau}_{t_0}\!\!\!\!\!\!\!\! d\tau_2u_L(t,\tau_1)[4g(\tau_2,\tau_1)]u_L(t+\tau,\tau_2),  \label{cgf}
\end{align}
see the derivation in Appendix B.

In Fig.~\ref{vL}, we present a 3D plot for the amplitude of the two-time correlation $v_L(t,t+\tau)$ at $T=0$. At operation time $t$, the pairing-induced vacuum quantum
fluctuation, characterized by $v_L(t,t)$, is built-up in time. Then the correlation function $v_L(t,t+\tau)$ generally decays as the delay time $\tau$ evolves, but never decay to zero.  This is again due to the dissipationless localized bound state discussed in the last section, i.e. the contribution from the first term in Eq.~(\ref{u_L}), so that the non-Markovian memory can be partially preserved at $T=0$. On the other hand, at $T\neq 0$, we find that the BCS vacuum quantum fluctuations with different operation times $t$ eventually decay to zero, indicating a complete memory less in the steady-state limit.
\begin{figure}
\centerline{\scalebox{0.26}{\includegraphics{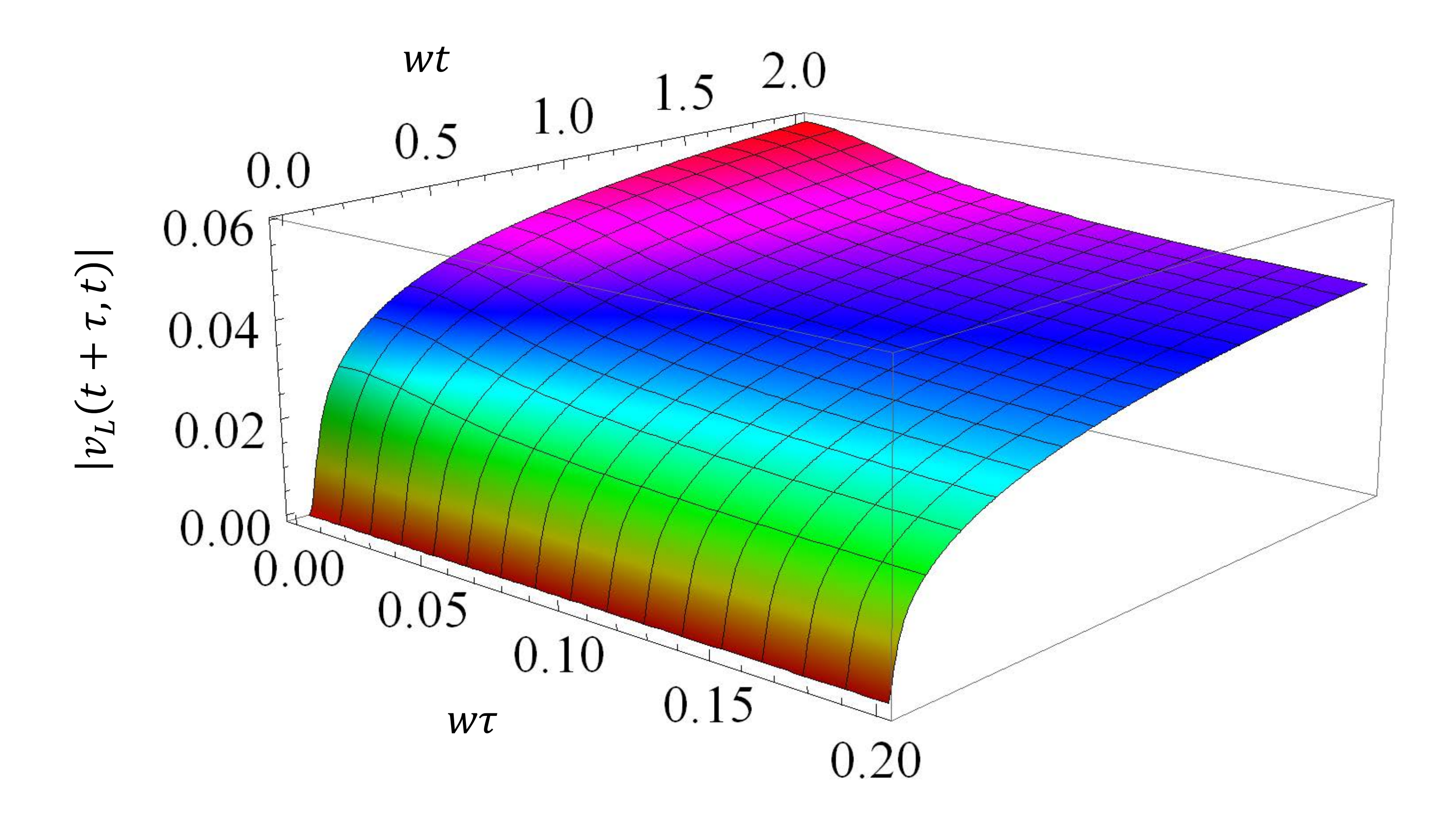}}}
\caption{(color online) The amplitude of the correlation function $v_L(t,t+\tau)$ as a function of the operation time $wt$ and the delay time $w\tau$, with the charge fluctuation strength $\frac{\pi}{2}B_G\eta^2=0.01$ at $T=0$. The gap parameter $\Delta=0.05w$.}\label{vL}
\end{figure}

\section{Conclusion}
We have derived the exact master equation of the zero-energy bogoliubon as well as Majorana zero modes that locally coupled to controlling gates. We have shown from this exact master equation that the decoherence of Majorana zero modes, induced by the charge fluctuations of the controlling gate, is actually a spin-like pure dephasing process. The dephasing rate is given by the decay rate of the left Majorana zero mode, which can be fully determined by the retarded Green function of the left Majorana mode. It should be pointed out the the exact master equation, Eqs.~(\ref{ME}) or (\ref{mME}) derived in this paper, is general, and can be applied to other topological systems with zero-energy bogoliubon or Majorana zero modes,
as long as the coupling Hamiltonian between the zero-energy bogoliubon or Majorana zero modes with other bogoliubons or other fermions is given by
\begin{align}
H_T &= \eta \sum_{k}V_{0k} [b^\dag_0a_{k}+b^\dag_0a^\dag_{k}]+H.c. \notag \\
&= -i\eta \sum_{k} V_{0k}\gamma_L(a_k+a^\dag_k)  .
\end{align}
Here $V_{0k}$ is the coupling amplitude of the Majorana zero mode with other fermion mode $k$. Charge fluctuations discussed in this paper correspond to $V_{0k}= \delta Q X_{0k}$.

Based on the resulting exact master equation, we analyzed in detail the non-Markovian decoherence dynamics of Majorana zero modes at $T=0$ as well as $T\neq 0$ of the controlling gate.
Physically, non-Markovian dynamics is defined as memory processes in open quantum systems. It is characterized by different-time correlations describing the environment-induced dissipation and fluctuation memory dynamics. Our results show that at zero temperature, there exists a zero-energy localized bound state for the left Majorana mode, which is a bound state resulted from the zero point in the spectral density at zero frequency. The existence of this localized bound state partially protects Majorana zero modes from decoherence at zero temperature, as a typical long-time non-Markovian memory effect. Also, as charge fluctuations get strong, the left Majorana mode decays nonexponentially such that the decay rate oscillates between positive and negative values in the short time regime, as a short-time non-Markovian memory effect. Thus, the decoherence dynamics of the left Majorana zero mode contains both the long-time and short-time non-Markovian memory effects at zero temperature.
At finite temperature, the zero-energy localized bound state becomes a resonant state with a very small linewidth arisen from the sharp dip of the spectral density near zero frequency. The corresponding decoherence dynamics of the left Majorana zero mode mainly manifests the short-time non-Markovian memory effect. The left Majorana mode will be eventually destroyed in the long time limit.


Furthermore, we studied the two-time correlation functions between two Majorana zero modes. The left-right Majorana mode correlations indicate that although only one of the Majorana zero modes is disturbed by charge fluctuations, the local-perturbation-induced decoherence may affect globally Majorana-mode based qubit operations. We also find that decoherence of Majorana zero modes cannot be fully
protected as increasing the gap parameter $\Delta$ between the zero and finite energy modes. The decoherence behavior for
the left Majorana zero mode at the ideal case of $\Delta = w$ given in Sec.~IIIB, where $w$ is the hopping
parameter in the Kitaev model, is indeed very similar to the case for a rather small gap parameter $\Delta=0.05w$
shown in Sec.~IIIA. The reason for such similar decoherence dynamics is due to the fact that when $\Delta \rightarrow w$, the coupling
$X_{0k}$ between the zero- and finite-energy bogoliubons, see Eq.~(\ref{XY}), becomes stronger accordingly.
As a result, the small and large gap parameters lead to similar decoherence dynamics for Majorana zero modes under
local charge fluctuations.

In conclusion, we find that decoherence to Majorana zero modes is inevitable under local perturbations,
even though it may be partially protected from localized bound states. The long-time and short-time
non-Markovian memory dynamics for Majorana zero modes intimately rely on the spectral structure and
temperature of the environment, as well as the coupling between the system and the environment.
They are manifested respectively by the existence of localized bound states and the nonexponential
decays with decay rate oscillating between positive and negative values. It reproduces the general
non-Markovian dynamics in open quantum systems [\onlinecite{PRL2012}].
Such a understanding of non-Markovian dynamics can help one to engineer
and control decoherence in practical applications of Majorana zero modes for quantum information processing.

\section*{Acknowledgement}
This work is supported by the Ministry of Science and Technology of the Republic of China under the contract No. MOST 105-2112-M-006-008-MY3.

\appendix
\section{Derivation of the exact master equation of the zero-energy bogoliubon}
In this appendix, we extend the Feynman-Vernon influence functional approach to fermionic coherent state representation incorporating pairing excitations to derive the exact master equation Eq.~(\ref{ME}) for the zero-energy bogoliubon. The zero-energy bogoliubon is disturbed by charge fluctuations induced from the controlling gate, which is described by Eq.~(\ref{HT}). In the fermionic coherent state representation with the initial state (\ref{rho initial}) of the total system, the reduced density matrix (\ref{trace rho}) at arbitrary time $t$ is expressed as [\onlinecite{PRB2008}, \onlinecite{NJP2010}]
\begin{align}
\langle\xi_t&|\rho(t)|\xi'_t\rangle \notag\\
&=\!\! \int \!\! d\mu(\xi_0)d\mu(\xi'_0)\langle\xi_0|\rho(t_0)|\xi'_0\rangle \mathcal{K}(\xi^*_t,\xi'_t,t|\xi_0,\xi'^{*}_0,t_0).
\end{align}
where $\rho(t_0)$ is an arbitrary initial state of the zero-energy bogoliubon. The variables $\xi^*_t,\xi'_t, \xi_0,\xi'^{*}_0 $
are Grassmann numbers, which are the eigenvalues of zero-energy bogoliubon operators $b^\dag_0, b_0$ in the fermionic
coherent states: $|\xi \rangle = e^{-\xi b^\dag_0}|0\rangle$ that $b_0 |\xi \rangle= \xi |\xi \rangle$ and $\langle \xi | b^\dag_0 = \xi^* \langle \xi |$  [\onlinecite{PRB2008}].
The propagating function $\mathcal{K}(\xi^*_t,\xi'_t,t|\xi_0,\xi'^{*}_0,t_0)$ is given by.
\begin{align}
\mathcal{K}&(\xi^*_t,\xi'_t,t|\xi_0,\xi'^{*}_0,t_0)\notag\\
&=\!\! \int \!\! \mathcal{D}[\xi^*,\xi,\xi'^{*},\xi']e^{i(S_S[\xi^*,\xi]-S^*_S[\xi'^{*},\xi'])}\mathcal{F}[\xi^*,\xi,\xi'^{*},\xi'],
\label{propagating function}
\end{align}
where
\begin{align}
S_S[\xi^*,\xi]&=-\frac{i}{2}[\xi^*_t\xi(t)+\xi^*(t_0)\xi_0]\notag\\
&-\frac{i}{2}\int^t_{t_0}d\tau[\dot{\xi}^*(\tau)\xi(\tau)-\xi^*(\tau)\dot{\xi}(\tau)] ,
\end{align}
is the action for the zero-energy bogoliubon. The influence functional $\mathcal{F}[\xi^*,\xi,\xi'^{*},\xi']$ is obtained after integrating over all the degrees of freedom of the finite-energy bogoliubons and also that of the controlling gate,
\begin{align}
\mathcal{F}[\xi^*,\xi,  \xi'^{*}, & \xi'] \notag \\
 =\exp\Bigg\{&\!\! - \!\! \int^t_{t_0} \!\! d\tau \!\! \int^\tau_{t_0} \!\! d\tau' \!\begin{pmatrix}\xi^*(\tau) & \xi(\tau)\end{pmatrix} \bm G(\tau,\tau')
\!\begin{pmatrix}\xi(\tau') \\ \xi^*(\tau')\end{pmatrix}\notag\\
&\!\! - \!\! \int^t_{t_0} \!\! d\tau \!\! \int^\tau_{t_0} \!\! d\tau' \!\begin{pmatrix}\xi'^{*}(\tau') & \xi'(\tau')\end{pmatrix}\bm G(\tau',\tau)
\!\begin{pmatrix}\xi^{'}(\tau) \\ \xi'^{*}(\tau)\end{pmatrix}\notag\\
&\!\! -\!\! \int^t_{t_0} \!\! d\tau\!\! \int^t_{t_0} \!\! d\tau' \!\begin{pmatrix}\xi'^{*}(\tau) & \xi'(\tau)\end{pmatrix} \bm G(\tau,\tau')
\!\begin{pmatrix}\xi(\tau') \\ \xi^*(\tau')\end{pmatrix}\!\! \Bigg\}, \label{inff}
\end{align}
where $\bm G(\tau,\tau')$ is given by Eq.~(\ref{Gtilde}). The necessity of using matrix formula here is due to the pairing excitations presented in the superconducting Hamiltonian.

Because the effective action [the action of the zero-energy bogoliubon plus the one obtained from the influence functional (\ref{inff})] is quadratic in terms of $\{\xi, \xi^*, \xi', \xi'^{*}\}$, the path integral of (\ref{propagating function}) can be exactly carried out using the stationary-phase approach. The resulting propagating function is
\begin{align}
\mathcal{K}&(\xi^*_t,\xi'_t,t|\xi_0,\xi'^{*}_0,t_0)\notag\\
=&N(t)\exp\Big\{\frac{1}{2}[\xi^*_t\xi(t)+\xi^*(t_0)\xi_0+\xi'^{*}(t)\xi'_t+\xi'^{*}_0\xi'(t_0)]\Big\},
\label{propagating stationary}
\end{align}
where $\xi(t), \xi^*(t_0), \xi'(t_0), \xi'^{*}(t)$ are determined by the stationary path which obey the following equations of motion,
\begin{subequations}\label{stationary path}
\begin{align}
\frac{d}{d\tau}
\begin{pmatrix}\xi(\tau)\\ \xi^{*}(\tau)\end{pmatrix}=- \! \int^\tau_{t_0} \!\! d\tau' \tilde{\bm G}(\tau,\tau')
\begin{pmatrix}\xi(\tau')\\ \xi^{*}(\tau')\end{pmatrix} ~~~~~ \notag\\
+ \! \int^t_{t_0} \!\! d\tau' {\bm G}(\tau',\tau)
\begin{pmatrix}\xi(\tau') + \xi'(\tau') \\ \xi^{*}(\tau')+\xi'^*(\tau') \end{pmatrix}, \\
\frac{d}{d\tau}
\begin{pmatrix}\xi'(\tau)\\ \xi'^*(\tau)\end{pmatrix}=- \! \int^\tau_{t_0} \!\! d\tau'  \tilde{\bm G} (\tau,\tau')
\begin{pmatrix}\xi'(\tau')\\ \xi'^*(\tau')\end{pmatrix} ~~~~~ \notag\\
 + \! \int^t_{t_0} \!\! d\tau'  {\bm G}(\tau,\tau')
\begin{pmatrix}\xi(\tau') + \xi'(\tau') \\ \xi^{*}(\tau')+\xi'^*(\tau')\end{pmatrix} .
\end{align}
\end{subequations}
By introducing the following transformation [\onlinecite{PRB2008}],
\begin{subequations}
\begin{align}
&\begin{pmatrix}\xi(\tau)\\ \xi^{*}(\tau)\end{pmatrix}=\bm{U}(\tau,t_0)\begin{pmatrix}\xi_0\\ \xi^{*}(t_0)\end{pmatrix}
+\bm{V}(\tau,t)\Big{[}\begin{pmatrix}\xi(t) + \xi'_t \\ \xi^{*}_t +\xi'^*(t)\end{pmatrix}\Big{]}\\
&\begin{pmatrix}\xi(\tau) + \xi'(\tau) \\ \xi'^{*}(\tau)+\xi^*(\tau)\end{pmatrix}
=\bm{U}^\dag(t,\tau)\begin{pmatrix}\xi(t) + \xi'_t \\ \xi^*_t+ \xi'^{*}(t) \end{pmatrix} ,
\end{align}
\end{subequations}
where $\bm{U}(\tau,t_0)$ and $\bm{V}(\tau,t)$ obey the integro-differential equations of motion (\ref{UV}), obtained from
Eq.~(\ref{stationary path}),
one can find that
\begin{widetext}
\begin{align}
\mathcal{K}(\xi^*_t,\xi'_t,t|\xi_0,\xi'^{*}_0,t_0)=N(t)\exp\Big\{&\begin{pmatrix}
\xi^*_t & \xi'_t
\end{pmatrix}\bm{J}_1(t)\begin{pmatrix}\xi_0\\ \xi'^{*}_0\end{pmatrix}+\begin{pmatrix}
\xi^*_t & \xi'_t
\end{pmatrix}\bm{J}_2(t)\begin{pmatrix}\xi'_t\\ \xi^*_t\end{pmatrix}\notag\\
&+\begin{pmatrix}
\xi'^{*}_0 & \xi_0\end{pmatrix}\bm{J}_3(t)\begin{pmatrix}\xi_0\\ \xi'^{*}_0\end{pmatrix}+\begin{pmatrix}
\xi'^{*}_0 & \xi_0
\end{pmatrix}\bm{J}^{\dag}_1(t)\begin{pmatrix}\xi'_t\\ \xi^*_t\end{pmatrix}\Big\},
\end{align}
and
\begin{subequations}
\label{Js}
\begin{align}
\bm{J}_1(t)&=\frac{1}{2}\bm{Z}\Big{[}\begin{pmatrix}1 & 0 \\0 & 0\end{pmatrix}-\bm{V}(t,t)-\bm{U}(t,t_0)\begin{pmatrix}0 & 0 \\0 & 1\end{pmatrix}\bm{U}^{\dag}(t,t_0)\Big{]}^{-1}\bm{U}(t,t_0)\bm{Z}\\
\bm{J}_2(t)&=\frac{1}{2}\bm{Z}\Big\{\Big{[}\begin{pmatrix}1 & 0 \\0 & 0\end{pmatrix}-\bm{V}(t,t)-\bm{U}(t,t_0)\begin{pmatrix}0 & 0 \\0 & 1\end{pmatrix}\bm{U}^{\dag}(t,t_0)\Big{]}^{-1}\bm{Z}-\bm{I}\Big\}\\
\bm{J}_3(t)&=\frac{1}{2}\bm{Z}\Big\{\bm{U}^{\dag}(t,t_0)\Big{[}\begin{pmatrix}1 & 0 \\0 & 0\end{pmatrix}-\bm{V}(t,t)-\bm{U}(t,t_0)\begin{pmatrix}0 & 0 \\0 & 1\end{pmatrix}\bm{U}^{\dag}(t,t_0)\Big{]}^{-1}\bm{U}(t,t_0)\bm{Z}-\bm{I}\Big\}.
\end{align}
\end{subequations}
\end{widetext}
Then, using the same differential form for the fermion creation and annihilation operators in the coherent state representation [\onlinecite{PRB2008}], we finally obtain the exact master equation for the zero-energy bogoliubon, given by Eq.~(\ref{ME}).

It should be pointed out that although the derivation of the master equation given above exactly follows the procedure of
[\onlinecite{PRB2008}], the derivation is much more difficult due to the existence of pairing coupling. Also, the new master
equation contains new dissipation processes induced by pair productions and annihilations.
Moreover, because the TSC is initially in Bogoliubov quasiparticle vacuum state, one can simplify further the above
time-dependent coefficients in the master equation. To do so, we may write the $2\times 2$ retarded Green function matrix
$\bm U (t,t_0)$ explicitly as,
\begin{align}
\bm U (t,t_0)= \begin{pmatrix}
u_{11}(t,t_0) & u_{12}(t,t_0)\\ & \\
u_{21}(t,t_0) & u_{22}(t,t_0) \end{pmatrix}
\end{align}
Then the equation of motion (\ref{UV}) for $\bm U (t,t_0)$ can be reexpressed in terms of its following components,
\begin{subequations}
\begin{align}
&\frac{d}{dt}[u_{11}(t,t_0)-u_{21}(t,t_0)]\label{U11mU12}\notag\\
&~~~~~=- 4 \!\! \int^t_{t_0} \!\!\! d\tau{\rm Re}[g(t,\tau)][u_{11}(\tau,t_0)-u_{21}(\tau,t_0)]\\
&\frac{d}{dt}[u_{11}(t,t_0)+u_{21}(t,t_0)]=0\label{U11pU12}\\
&\frac{d}{dt}[u_{22}(t,t_0)-u_{12}(t,t_0)] \notag \\
&~~~~~=- 4 \!\! \int^t_{t_0} \!\!\! d\tau {\rm Re}[g(t,\tau)][u_{22}(\tau,t_0)-u_{12}(\tau,t_0)] \label{U22mU21}  \\
&\frac{d}{dt}[u_{22}(t,t_0)+u_{12}(t,t_0)]=0,\label{U22pU21}
\end{align}
\end{subequations}
with the initial conditions $u_{11}(t_0,t_0) = u_{22}(t_0,t_0)=1$ and $u_{12}(t_0,t_0)=u_{21}(t_0,t_0)=0$.
Using these initial conditions, one can find from Eq.~(\ref{U11mU12}) and (\ref{U22mU21}) that
\begin{align}
u_{11}(t,t_0)-u_{21}(t,t_0)=u_{22}(t,t_0)-u_{12}(t,t_0),
\end{align}
and Eq.~(\ref{U11pU12}) and (\ref{U22pU21}) lead to
\begin{align}
u_{11}(t,t_0)+u_{21}(t,t_0)=u_{22}(t,t_0)+u_{12}(t,t_0).
\end{align}
As a result, we obtain the identities:
\begin{align}
u_{11}(t,t_0)=u_{22}(t,t_0), ~~
u_{12}(t,t_0)=u_{21}(t,t_0).
\end{align}
Furthermore, one can find that $\bm{V}_{11}(\tau,t)=\bm{V}_{22}(\tau,t)=-\bm{V}_{12}(\tau,t)=-\bm{V}_{21}(\tau,t)$.

Also, from the equation of motion (\ref{UV}), we have
\begin{align}
\frac{d}{d\tau_1}\Big[\bm{U} & (\tau,\tau_1)\bm{U}^{\dag}(t,\tau_1)\Big]\notag\\
&=\int_{\tau_1}^{\tau}d\tau_2\bm{U}(\tau,\tau_2)\bm{G}(\tau_2,\tau_1)\bm{U}^{\dag}(t,\tau_1)\notag\\
&~~~+\int_{\tau_1}^{t}d\tau_2\bm{U}(\tau,\tau_1)\bm{G}(\tau_1,\tau_2)\bm{U}^{\dag}(t,\tau_2).
\end{align}
Integrating the both sides of the above relation, one can find that [\onlinecite{Lo-pri}]
\begin{align}
\bm{U}^{\dag} & (t,\tau)-\bm{U}(\tau,t_0)\bm{U}^{\dag}(t,t_0)\notag\\
&=\int_{t_0}^{\tau}d\tau_1\int_{t_0}^{t}d\tau_2\bm{U}(\tau,\tau_1)\bm{G}(\tau_1,\tau_2)\bm{U}^{\dag}(t,\tau_2).\label{U identity}
\end{align}
Using the fact that $g(\tau_1,\tau_2)=g(\tau_1-\tau_2)$ [see Eq.~(\ref{singg})] and
$\bm{U}(t,t_0)=\bm{U}^{\dag}(t,t_0)$, we have the following identity 
\begin{align}
\bm{V}(\tau,t)&+\bm{V}(t,\tau)=2{\rm{Re}}[\bm{V}(\tau,t)]\notag\\
&=\bm{U}(t,\tau)-\bm{U}(\tau,t_0)\bm{U}(t,t_0).
\label{V identity}
\end{align}
Using this identity, we can prove that the time-dependent dissipation and fluctuation coefficients in the master
equation (\ref{ME}) are all related to each other,
\begin{align}
\lambda(t,t_0)=-\Lambda(t,t_0)=\widetilde{\lambda}(t,t_0).
\end{align}
We can also express the matrix elements of $\bm{U}(t,t_0)$ in terms of the retarded Green functions
$u_L(t,t_0)$ and $u_R(t,t_0)$ (notice that $u_R(t,t_0)=1$). Then
$\lambda(t,t_0)$ is simply given by,
\begin{align}
\lambda(t,t_0)=-\frac{1}{2}\frac{\dot{u_L}(t,t_0)}{u_L(t,t_0)}.
\end{align}
This leads to the simple master equation for the Majorana zero mode that is
equivalent to a spin-like pure dephasing model, see Eq.~(\ref{mME}).

\section{Equation of motion for Majorana zero modes}

From the definition of the Majorana operators Eq.~(\ref{gamma}),
\begin{align}
\gamma_L=-i(b_0-b^{\dag}_0),  ~~\gamma_R=b_0+b^{\dag}_0,
\end{align}
it is easy to find the corresponding Heisenberg equations of motion,
\begin{align}
&\frac{d}{dt}\!\! \begin{pmatrix} \gamma_L(t) \\ \gamma_R (t) \end {pmatrix}= \begin{pmatrix} - 2\eta\delta Q(t)\sum_kX_{0k}
[b_k(t)+b^\dag_k(t)] \\ 0 \end{pmatrix} .
\label{ege}
\end{align}
Also, using the formal solution of $b_k(t)$ and $b^\dag_k(t)$ from their Heisenberg equations of motion, we obtain further that
\begin{subequations}
\label{mfem}
\begin{align}
\frac{d}{dt} \gamma_L(t) =&-4\! \int^t_{t_0} \!\!\! d\tau {\rm Re}[g(t,\tau)]\gamma_L(\tau) + \Gamma_L(t) , \\
 \frac{d}{dt} \gamma_R(t) = & 0.
\end{align}
\end{subequations}
where
\begin{align}
\Gamma_L(t)=- 2\eta\delta Q(t) \! \sum_k X_{0k} \big[ b_k(t_0)e^{-iE_k(t-t_0)} + {\rm H.c}\big]
\end{align}
is the environment-induced noise force operator.
Eq.~(\ref{mfem}) shows that the right Majorana mode is explicitly decoupled from the environment. The left Majorana mode
obeys a linear differential-integral equation (a generalized quantum Langevin equation [\onlinecite{PRB2015}]), its formal solution can be
expressed as
\begin{align}
\gamma_L(t) = u_L(t, t_0)\gamma_L(t_0) + f_L(t),
\end{align}
where $u_L(t,t_0)$ is the retarded Green function defined as $u_L(t,t_0) = \frac{1}{2}\langle \{\gamma_L(t),\gamma_L(t_0)\} \rangle$ which
satisfies a simple Kadanoff-Baym type equation in the nonequilibrium Green function formalism,
\begin{align}
\frac{d}{dt}u_L(t,t_0)= - 4 \! \int^t_{t_0} \!\!\! d\tau  {\rm Re}[g(t,\tau)] u_L(\tau,t_0) ,
\end{align}
subjected to the initial condition $u_L(t_0,t_0)=1$.
The operator $f_L(t)$ is the noise-force induced fluctuation field, the corresponding equation of motion is given by
\begin{align}
\frac{d}{dt}f_L(t) = - 4 \! \int^t_{t_0} \!\!\! d\tau  {\rm Re}[g(t,\tau)] f_L(\tau,t_0) + \Gamma_L(t) ,
\end{align}
with the initial condition $f_L(t_0)=0$. The solution of the above equation is
\begin{align}
f_L(t)= \int_{t_0}^t \!\!\! d \tau u_L(t,\tau) \Gamma_L(\tau).
\end{align}
Then the two-time correlation function of the left Majorana mode is given by
\begin{align}
\langle \gamma_L(t+\tau) \gamma_L(t) \rangle &= u_L(t+\tau, t_0)u_L(t,t_0) \notag\\
&+ \langle f_L(t+\tau) f_L(t) \rangle   \label{LLc}
\end{align}
and
\begin{align}
\langle f_L(t) f_L(t') \rangle &= \!\! \int^{t}_{t_0} \!\!\! d\tau \!\! \int^{t'}_{t_0} \!\!\! d\tau' u_L(t, \tau)\widetilde{g}_{LL}(\tau, \tau') u_L (t',\tau') \notag\\
&\equiv v_L(t',t).  \label{lfc}
\end{align}
To be more specific, the correlation function $\widetilde{g}_{LL}(\tau,\tau')$ can be explicitly written as
\begin{align}
\widetilde{g}_{LL}&(\tau,\tau')\notag\\
&=4\eta^2\sum_k X^2_{0k}\langle\delta Q(\tau)\delta Q(\tau')\rangle\langle b^{\dag}_k(t_0)b_k(t_0)\rangle e^{-iE_k(\tau-\tau')}\notag\\
&+4\eta^2\sum_k X^2_{0k}\langle\delta Q(\tau')\delta Q(\tau)\rangle\langle b_k(t_0)b^{\dag}_k(t_0)\rangle e^{-iE_k(\tau'-\tau)}
\end{align}
As the TSC is initially in the vacuum bogoliubon state, namely, $\langle b^{\dag}_k(t_0)b_k(t_0)\rangle=0$, the correlation function $\widetilde{g}_{LL}(\tau',\tau)$ describes the pure vacuum quantum fluctuation induced by the non-zero pair excitation. Therefore,
\begin{align}
\widetilde{g}_{LL}(\tau,\tau')&=4\eta^2\sum_k X^2_{0k}\langle\delta Q(\tau')\delta Q(\tau)\rangle e^{-iE_k(\tau'-\tau)}\notag\\
&=4g(\tau',\tau)
\end{align}
Finally, the above result can be written simply as
\begin{align}
\langle \gamma_L(t+\tau) \gamma_L(t) \rangle = u_L(t+\tau, t_0)u_L(t,t_0) + v_L(t,t+\tau)
\end{align}

\end{document}